\documentclass[12pt,a4paper]{article}
\usepackage{type1cm,amsmath,hangcaption,indentfirst}
\usepackage[psamsfonts]{amssymb}
\usepackage{amsfonts}
\usepackage{mathrsfs}
\usepackage{latexsym}
\usepackage{graphics}
\usepackage{sectsty}
\usepackage{exscale,latexsym}
\usepackage[dvips]{graphicx}
\usepackage{setspace}
\usepackage{multirow}
\usepackage{epsfig}
\usepackage{graphics}
\usepackage{graphicx,indentfirst}
\usepackage{psfrag}
\numberwithin{equation}{section}

\newcommand{\half}{1\over 2}

\newcommand{\C}{\mathbb{C}}

\newcommand{\del}{\partial}

\sectionfont{\normalsize\bf}
\subsectionfont{\rm\normalsize}
\def\be{\begin{equation}}
\def\ee{\end{equation}}
\def\half{{\scriptstyle {1\over 2}}}

\def\quarter{{\scriptstyle {1\over  4}}}

\newcommand{\da}{\dot{a}}
\newcommand{\db}{\dot{b}}
\newcommand{\bpi}{\bar{\pi}}

\def\C{{\cal C}}
\def\M{{\cal M}}

\def\Str{{\rm Str}}

\def\boxit#1{\vbox{\hrule\hbox{\vrule\kern5pt
	 \vbox{\kern5pt#1\kern5pt}\kern5pt\vrule}\hrule}} 
\def\vac{|0\rangle}

\def\O{{\cal O}}

\def\M{{\cal M}}

\def\zb{{\bar z}}

\def\bpi{{\overline\pi}}

\def\noblackbox{\overfullrule=0pt}
\noblackbox
\def\half{{\scriptstyle {1\over 2}}}

\def\quarter{{\scriptstyle {1\over  4}}}

\def\Zzb{\overline Z}
\def\Zz{Z}

\begin{document}
%%% Title page %%%%%
\begin{titlepage}
\renewcommand{\thefootnote}{\fnsymbol{footnote}}
\begin{flushright}
\begin{tabular}{l}
arXiv:1008.xxxx\\ %This should be replaced after submission.
\end{tabular}
\end{flushright}

% \vfill
\begin{center}

\vskip 2.5 truecm

\noindent{\large \textbf{Yangian in the Twistor String}}\\
\vspace{1.5cm}

\noindent{ John Corn, %\footnote{E-mail:}, 
Thomas Creutzig %\footnote{E-mail: creutzig@physics.unc.edu} 
and Louise Dolan%\footnote{E-mail:}}
}\bigskip

 \vskip .6 truecm
\centerline{\it Department of Physics and Astronomy}
\centerline{\it University of North Carolina, Chapel Hill, NC 27599, USA}

 \vskip .4 truecm

 \end{center}

 \vfill
\vskip 0.5 truecm

\begin{abstract}
We study symmetries of the quantized open twistor string. 
In addition to global PSL(4$|$4) symmetry, we find non-local 
conserved currents.
The associated non-local charges lead to Ward identities which show that
these charges annihilate the string gluon tree amplitudes, and have
the same form as symmetries of amplitudes in $\mathcal N=4$ super
conformal Yang Mills theory. We describe how states of the open twistor 
string form a realization of the PSL(4$|$4) Yangian superalgebra.

\end{abstract}
\vfill
\vskip 0.5 truecm

\setcounter{footnote}{0}
\renewcommand{\thefootnote}{\arabic{footnote}}
\noindent{E-mail:  jrcorn@physics.unc.edu, \;creutzig@physics.unc.edu,
\; ldolan@physics.unc.edu}

\end{titlepage}

\newpage

\tableofcontents
%%%%%%%%%%%%%%%%%%%%%%%%%%%%%%%%%%%%%%%%%%%%%%%%%%%%%%%%%%%%%%%%%%%%%%
%\newpage

\vfill\eject

\section{Introduction}
\label{sec:introduction}

Twistor string theory \cite{W}-\cite{BW}
is equivalent to a massless field theory and
provides a string structure to analyze four-dimensional 
massless scattering amplitudes with $N=4$ supersymmetry.
At tree level, the spectrum can describe states of both super Yang Mills 
theory and conformal supergravity. The world sheet theory has 
a target in super twistor space, and has been argued to have
a Yangian extension of its $PSL(4|4)$ global symmetry \cite{W}. 

In this paper we construct non-local conserved currents which lead
to a Hopf algebra coproduct, show the associated charges
annihilate the string gluon tree amplitudes and give a realization of the
Yangian symmetry algebra for this world sheet action.
We leave the conformal graviton amplitudes for future analysis.
Explicit constructions of the 
amplitudes have revealed some of the string's framework
\cite{DG3}-\cite{RMV3}.
How the Yangian acts on the various building blocks of the theory
\cite{ACC}-\cite{ABCT2} 
should help further unravel the theory, at least in the planar
limit. The string also relates to other twistor methods
\cite{Mason}-\cite{Wolf}.  

The Yangian symmetry is expected
because of the string's close connection to Yang Mills theory,
for which an infinite-dimensional symmetry for the planar theory
has been displayed 
\cite{DNW1}-\cite{pop}, 
and its appearance as dual conformal invariance was found explicitly
for the amplitudes \cite{Drummond1}-\cite{DF2}.

We review Yangians and their nature in field theory.
A Yangian superalgebra is a superalgebra, that is an algebra that has 
bosonic and fermionic elements,  and it is graded by the
non-negative integers. The subalgebra of level zero is a Lie superalgebra and 
the grade one piece is its adjoint representation. The higher level parts 
of the superalgebra are generated by the algebra multiplication of elements 
of level one.  They are subject to some restrictions which arise from the 
comultiplication. The comultiplication is an important additional structure 
that specifies how the elements of the Yangian superalgebra act on products 
of states.  For an ordinary Lie superalgebra this is trivial
\begin{equation}\label{eq:trivialcoproduct}Q_0(A\,B) \ = 
\ Q_0(A)\,B+(-1)^{|Q_0||A|}A\,Q_0(B)\, .
\end{equation}
But the elements of level one of the Yangian have non-trivial 
comultiplication. They act schematically as
\begin{equation}\label{eq:nontrivialcoproduct}Q_1(A\,B) \ 
= \ Q_1(A)\,B+(-1)^{|Q_1||A|}A\,Q_1(B) + 
Q_0(A)\,Q_0(B)\, .\end{equation}
Now one needs to require that this operation is compatible with the algebra 
product. This can be ensured by the Serre relations. 
The question is how symmetries with such non-trivial action on products of 
states can arise in field theory.  For a model specified by a 
Lagrangian plus a path integral measure, with both invariant under some 
global symmetry, usually 
we have corresponding conserved {\em local} Noether currents 
and charges. But the action of charges on products of states is then of the 
form (\ref{eq:trivialcoproduct}).
The way to realize a non-trivial action on products of fields is by 
introducing {\em non-local} conserved charges. Non-locality is then 
reflected in the comultiplication 
\eqref{eq:nontrivialcoproduct}.

The world-sheet theory of the open twistor string is a two-dimensional 
conformal field theory, that is a local model. Introducing non-local fields 
thus seems strange. 
Nonetheless, in \cite{Bernard2}, it was shown how Yangian 
symmetry arises via non-local currents in Lie group Wess-Zumino-Witten models.
We will follow their analysis in spirit, but we will use the non-local currents 
only as an intermediate step. We can do that due to the special nature of the 
global symmetry supergroup PSL(4$|$4) of the twistor string. Our strategy is as 
follows. We start by introducing non-local currents as in \cite{Bernard2},
\cite{DB}. 
We show that the associated charges satisfy the comultiplication of the 
level one elements of the psl(4$|$4) Yangian. 
Then we derive Ward identities for these 
charges. They simplify considerably due to the vanishing of the Killing 
form of psl(4$|$4). These Ward identities lead us to define new 
operators acting on the fields of the twistor string. These operators form a 
realization of the Yangian. In addition they form a symmetry because of the 
Ward identities we found, and the new level one charges
do not have the properties of the old charges, namely construction 
from a non-local current.

In section 2 we review the twistor string, describe the
ordinary $PSL(4|4)$ Noether currents and discuss their canonical 
quantization in the boundary conformal field theory. 
In section 3 we give the action of these ordinary currents
and their charges on the fields and the tree amplitudes.
The non-local Noether currents are described in section 4,
and their action on the fields and the string tree amplitudes
is shown in section 5. In section 6, we discuss the
Serre relations and the representation of the Yangian of $PSL(4|4)$.
In the appendices we review superalgebras and give 
the $PSL(4|4)$ structure constants and their properties, 
including the vanishing of the adjoint quadratic
Casimir (the Killing form).

\section{$PSL(4|4)$ Noether currents of the open twistor string}
\label{sec:ws}

The open twistor string can be described by the action
\begin{equation}
 S\ = \ S_{YZ}+S_G+S_{\text{ghost}}\, ,
\label{wsact}\end{equation}
with $S_G$ given by a conformal field theory of central charge $c=28$,
$S_{\text{ghost}}$ is the standard $c=-26$ ghost system of the 
bosonic string,  and
$S_{YZ}$ is the world sheet action for fields with a target 
of twistor superspace,
\be S_{YZ}
= \int i\left [Y^{z I} D_z \Zzb_I + Y^{\zb I} D_\zb \Zz_I \right]\;
g^\half d^2x\, \label{yzact}\ee
where 
$1\le I\le 8$,
$D_\alpha = \partial_\alpha -i A_\alpha$, and 
$g$ is the determinant of the world sheet metric with
Euclidean signature, $z=x_1+ix_2$, $\zb =x_1-ix_2$. 
The world sheet can be described as the upper half-plane, and thus 
has a boundary.

The equations of motion are
%\be D_\zb \Zz _ I = D_z \Zzb_I = 0,\qquad D_z'Y^{z I} = D'_\zb Y^{\zb I} =0,
%\qquad\hbox{for\,  $D'_\alpha=(\partial_\alpha + iA_\alpha)g^\half$}, 
%\nonumber\ee
\begin{align} &(\partial_\zb - i A_\zb) Z_I = 0,\qquad
(\partial_z - i A_z) \Zzb_I = 0, \qquad
 Y^{z I}_{; z} + i A_z Y^{z I} = 0,\qquad
%= \left (\partial_z + \gamma^{-1}\partial_z\gamma + i A_z \right)
%Y^{z I} = 0,\qquad
Y^{\zb I}_{; \zb} + i A_\zb Y^{\zb I} =0.
%= \left (\partial_\zb + \gamma^{-1}\partial_\zb\gamma + i A_\zb \right)
%Y^{\zb I} = 0.
&\label{eom}
\end{align}
The constraints from varying the world sheet gauge fields are
$$ Y^{\zb I} \Zz_I = Y^{z I}\Zzb_I =0,$$
and the boundary conditions in upper half-plane world sheet coordinates are
\begin{align}
\Zzb_I = U Z_I,\qquad Y^{z I} =U^{-1}Y^{\zb I},
\label{bc}\end{align}
where $U=e^{2i\alpha}$, 
for some function $\alpha$, which varies and is real on the boundary,
and is continuous up to multiples of $\pi$.  
The reality conditions are 
$\overline{Y^{z\, I}} = - Y^{\bar z\, I}$ for bosonic components
($1\le I\le 4)$, 
and $\overline{Y^{z,\,I}} = Y^{\bar z\,I}$ for fermionic  components
($5\le I\le 8$), 
and $\overline{A_z}=-A_\zb$. 

The twistor string has global PSL(4$|$4) symmetry. 
To write the 
Noether currents for PSL(4$|$4), we consider the following
symmetry transformations of the world sheet action (\ref{wsact}),
\begin{align}
Z_I \rightarrow e^{\rho_a T^{aJ}_I} Z_J,
\qquad \Zzb_I \rightarrow e^{\rho_a T^{aJ}_I} \Zzb_J,\qquad\cr
Y^{\zb I} \rightarrow Y^{\zb J} e^{-\rho_a T^{a I}_J} ,\qquad
Y^{zI} \rightarrow Y^{zJ} e^{-\rho_a T^{a I}_J},
\label{symtran}\end{align}
where the generators of the Grassman envelope of the superalgebra 
psl(4$|$4),  $\rho_aT^{aJ}_I$, are real and have zero
trace and zero supertrace.
The other fields in (\ref{wsact}) are singlets of psl(4$|$4)
and ultimately singlets of the Yangian, so
from now on we will discuss symmetries of (\ref{yzact}). 
See appendix \ref{app:super} for some background on the
supergroup PSL(4$|$4) and superalgebra psl(4$|$4). 
The infinitesimal transformations are 
\begin{align}
\delta \Zz_I = \rho_a T^{aJ}_I Z_J,\qquad
\delta \Zzb_I = \rho_a T^{aJ}_I \Zzb_J,
\label{ordsym}\end{align}
leaving the Lagrangian invariant.
The Noether current is
\begin{align}
J^z = g^\half j^z,\qquad J^\zb = g^\half j^\zb, \qquad \hbox{with}
\quad
 j^z = i Y^{zI} \rho_a T^{aJ}_I \Zzb_J,\qquad
j^\zb = i Y^{\zb I}  \rho_a T^{aJ}_I Z_J,
 \label{onc}\end{align}
where $\partial_\alpha J^\alpha = 0$
and $j^\alpha$ satisfies 
$j^\alpha_{\hskip3pt ;\alpha} = 0$ with use of the equations of 
motion (\ref{eom}), and also separately
$j^z_{\hskip3pt ;z }= 0$ and  $j^\zb_{\hskip3pt ;\zb} =  0$.
So we also have  
\be d\, j = 0.\label{extd}\ee

In addition to (\ref{symtran}),
the identity matrix $T_I^{\hskip3pt J} = \delta_I^{\hskip3pt J}$ 
and the $U(1)$ R-symmetry matrix $T_I^{\hskip3pt J} = 
(-1)^{\hbox{deg}\, I} \, \delta_I^{\hskip3pt J}$ also generate 
symmetries of the classical action. Here 
$\hbox{deg}\, I = 0$ when $I$ is a bosonic index, and
$\hbox{deg}\, I = 1$ when $I$ is a fermionic index.
For simplicity we will denote the degree by
the index, $(-1)^{\hbox{deg}\, I} \equiv (-1)^I$.
The transformation associated with the identity matrix
obeys the boundary condition (\ref{bc})
for any complex function $\rho$, and generates the $gl(1,\Bbb C)$ 
gauge invariance
which we use to choose $A_z = A_\zb = 0$. The $U(1)$ R-symmetry
corresponds to a current that has a conformal anomaly
and is not conserved in the quantum theory.

We review the canonical quantization \cite{DG1} of the conformal field 
theory (\ref{yzact}).
The commutation relations are found from  
\be Z_I(z) Y^J(\zeta) = :Z_I(z) Y^J(\zeta): + {\delta_I^J\over z-\zeta},
\label{ope}\ee
for $|z|>|\zeta|$, 
and where $Y^J \equiv g^\half Y^{\zb J}$ is holomorphic in the 
gauge where $A_z = A_\zb =0$.
The Virasoro current relevant for the twistor fields is given by 
\be L_{YZ}(z) = -\sum_J Y^J(z)\partial Z_J(z)\label{VA}\ee
and its operator product expansion with the chiral currents
$:Y^I(z)Z_J(z):$ is 
\begin{align} &L_{YZ} (z)  :Y^I(\zeta) Z_J(\zeta): \, \sim\, 
- \frac{\delta^I_J \;(-1)^{\hbox{deg}\, I}}{(z-\zeta)^{3}}
+ \frac{:Y^I(\zeta) Z_J(\zeta):}{(z-\zeta)^{2}}  
+ \frac{\partial:Y^I(\zeta) Z_J(\zeta):}{(z-\zeta)}\, .
\label{ope1}\end{align}
Here $\sim$ denotes equality up to the regular part. 
The anomalous term vanishes for the supertraceless currents, 
that is to say the psl(4$|$4) currents
and the $gl(1,\Bbb C)$ gauge current, so they are true primary fields
and are conserved in the quantum theory.

In order to define Noether charges, 
we consider the boundary theory explicitly.
In a boundary conformal field theory one specifies how fields 
are identified at the boundary, such as in (\ref{bc}),
\begin{align} -(-1)^I \bar Y^I = U^{-1} Y,\qquad
\bar Z = U Z.\label{bcc}\end{align}
Our theory possesses a current superalgebra symmetry in addition to 
Virasoro invariance (see, e.g., 
\cite{Creutzig:2009zz}-\cite{Creutzig:2007jy}.)

In such a situation one usually has boundary conditions of the form
\begin{equation}
J(z) \ = \ \Omega\bar J(\bar z)\qquad\text{for}\ z \ = \ \bar z\, .
\end{equation}
Here $\Omega$ is a map on the space of fields specifing the 
boundary conditions.
If this map is an automorphsim of the Lie superalgebra, 
then these gluing conditions preserve
the current algebra symmetry. In this case,
we can analytically continue the current $J(z)$ on the entire plane 
\begin{equation}
\begin{split}
J(z)\ = \ \Bigl\{ \begin{array}{ccc}
\ J(z)&\qquad z \ \text{in upper half plane}     \\
\ \Omega\bar J(z)&\qquad z \ \text{in lower half plane}      \\
\end{array} \ 
\end{split}
\end{equation}
Then the charge is defined by an integral that runs from boundary (a) to
boundary (b) of the open string,
where \begin{align}
Q = \int d\sigma j^0 = \epsilon_{\mu\nu}\int d x^\mu j^\nu
= \int_a^b  dz j^{\bar z} - \int_a^b d\bar z j^z
= \oint dz j^{\bar z},
\nonumber\end{align}
and we include a factor of ${1\over 2\pi i}$ in
$\oint$. See Figure 1.

\vskip10pt

\begin{figure}[htb!]
\label{fig:typical}
\centering%
\psfrag{J}{$J(z)$}
\psfrag{barJ}{$\Omega\bar J(\bar z)$}
\psfrag{a}{a}
\psfrag{b}{b}
\psfrag{lhp}{}
\psfrag{uhp}{}
\psfrag{z}{\fbox{$z$-plane}}
\centering
\includegraphics[width=8cm]{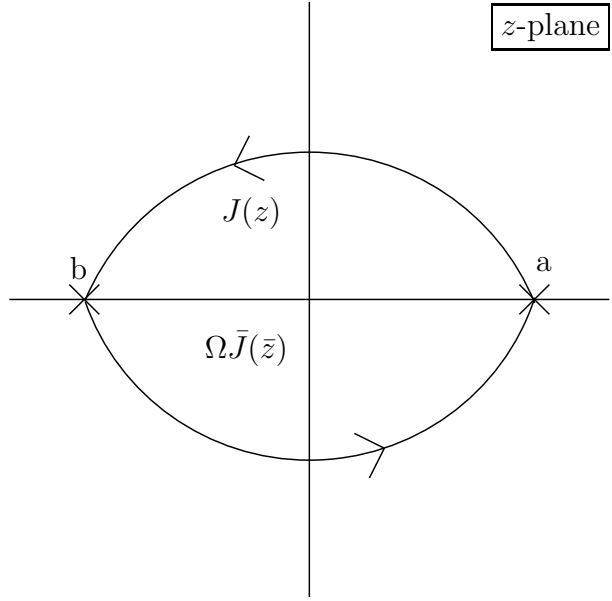}
\caption{\em Continuity of currents.}
\end{figure}

\vskip10pt
In our case, $\Omega = -(-1)^{IJ+I}$ which is an automorphism
of psl(4$|$4), and the conserved Noether charges associated with the 
transformations (\ref{ordsym}) can be written as
\begin{align}
Q^I_{0J} = \oint dz J^I_{0J}(z)
\label{noech}\end{align}
where the contour encirles the origin, and
the Nother currents for the psl(4$|$4) charges are
\begin{align}
J^I_{0\hskip2pt J}(z) \equiv Y^I(z)Z_J(z) - {1\over 8}(-1)^{I+E}\delta^I_J\,
Y^E(z)Z_E(z) - {1\over 8} \delta^I_J\,Y^E(z)Z_E(z).
\label{noecu}\end{align}
They have zero trace and zero supertrace and,
from (\ref{ope}), the current algebra satisfies
\begin{align}
J^I_{0\hskip2pt J}(z) \, J^K_{0\hskip2pt L}(\zeta) 
&\sim  (z-\zeta)^{-1} \, \Big(
f^{I\hskip3pt K\hskip14pt N}_{\hskip3pt J\hskip6pt L\,M}
\, J^M_{0\hskip2pt N}(\zeta)
+ {1\over 8}\Big( 1 - (-1)^{I+J}\Big) \delta^I_L\, \delta^K_J\;\;
Y^E(\zeta)Z_E(\zeta) \Big)\cr
&\hskip10pt  + (z-\zeta)^{-2} \left(- (-1)^I \delta^I_L \delta^K_J
+ {1\over 8} ( (-1)^I + (-1)^K) \delta^I_J \delta^K_L\right),
\label{Jca}\end{align}
where the psl(4$|$4) structure constants 
$f^{I\hskip3pt K\hskip14pt N}_{\hskip3pt J\hskip6pt L\,M}$
have vanishing Killing form
\begin{align}
f^{M\hskip6pt L\hskip6pt J}_{\hskip5ptN K\hskip5pt I}
\;\; f^{I\hskip3pt K\hskip14pt S}_{\hskip3pt J\hskip6pt L\,R}
&= 0,
\label{vkf}\end{align}
and are given in appendix \ref{app:structure}, together
with some of their properties.
The charges (\ref{noech}) satisfy
\begin{align}
[Q^I_{0\hskip2pt J}, \,Q^K_{0\hskip2pt L}\} = 
f^{I\hskip3pt K\hskip12pt N}_{\hskip3pt J\hskip6ptL\;M}\;
Q^M_{0\hskip2pt N} + 
{1\over 8}\Big( 1 - (-1)^{I+J}\Big) \delta^I_L\, \delta^K_J\;\;
\oint dz Y^E(z)Z_E(z)
\label{cha}\end{align}
Since the central generator $\oint dz Y^E(z)Z_E(z)$ acts as zero
on gauge invariant quantities like the vertex operators, 
the twistor string has psl(4$|$4) global symmetry, and
\begin{align}
[Q^I_{0\hskip2pt J}, \,Q^K_{0\hskip2pt L}\} =
f^{I\hskip3pt K\hskip12pt N}_{\hskip3pt J\hskip6ptL\;M}\;
Q^M_{0\hskip2pt N}\label{chainv}\end{align}
on gauge invariant states.
The mixed brackets in (\ref{cha})
denote an (anti) commutator for (odd) even 
generators,
\begin{align} [Q^I_{0\hskip2pt J}, \,Q^K_{0\hskip2pt L}\}\equiv
Q^I_{0\hskip2pt J}\, Q^K_{0\hskip2pt L} - (-1)^{(I+J)(K+L)}
\, Q^K_{0\hskip2pt L} Q^K_{0\hskip2pt L}.
\end{align}
See appendix \ref{app:realform} for a familiar basis.

\section{How the PSL(4$|$4) symmetry acts}
\label{sec:levelzero}

\subsection{\it Action of PSL(4$|$4) Currents and Charges on Fields}

The ordinary charges generate infinitesimal transformation on the
fields $Y^K(z)$ and $Z_K(z)$ and act on products
of more general conformal fields via standard Lie algebra
comultiplication, as follows. 

The operator product of the current $J^I_{0\hskip 2pt J} (z)$
with fields $Z_K(\zeta)$ is
\begin{align}
J^I_{0\hskip 2pt J} (z) \; Z_K(\zeta) &\sim
(z-\zeta)^{-1} \left( -(-1)^{IJ+K} \delta^I_K Z_J(z)
+ {1\over 8} (-1)^{I} \delta^I_J Z_K(\zeta)
+ {1\over 8} \delta^I_J Z_K(\zeta) \right)\cr
&\equiv (z-\zeta)^{-1} \; t^{\hskip2pt I}_{\hskip3pt J}
[\zeta] \; Z_K(\zeta),
\label{jzope}\end{align}
for $|z|>|\zeta|$ where we have defined the differential operations 
\begin{align} 
t^{\hskip2pt I}_{\hskip2pt J}[\zeta] &\equiv
-(-1)^{IJ+I} \,  Z_J(\zeta) {\partial\over\partial Z_I(\zeta)}
+{1\over 8} \delta^I_J (-1)^{I+E} Z_E(\zeta){\partial\over
\partial Z_E(\zeta)} 
+{1\over 8} \delta^I_J Z_E(\zeta){\partial\over
\partial Z_E(\zeta)}\cr
&\hskip15pt  + Y^I(\zeta) {\partial\over\partial Y^J(\zeta)}
- {1\over 8} \delta^I_J (-1)^{I+E} Y^E(\zeta){\partial\over
\partial Y^E(\zeta)} 
-{1\over 8} \delta^I_J Y^E(\zeta){\partial\over\partial Y^E(\zeta)}.
\label{tdef}\end{align}

Since $J^I_{0\,J}(z)$ and $Z_K(\zeta)$ are respectively local, 
we find using the standard contour argument, 
\begin{align}
[Q^I_{0\hskip2pt J}, Z_K(\zeta)\} &=
\oint_{\Gamma^>_0} dz\,
(z-\zeta)^{-1} t^I_J[\zeta] Z_K(\zeta)
- \oint_{\Gamma^<_0} dz\,
(z-\zeta)^{-1} t^I_J[\zeta] Z_K(\zeta)\cr
&= \oint_{\Gamma_\zeta} dz\,  t^I_J[\zeta] Z_K(\zeta)
=  t^I_J[\zeta] Z_K(\zeta),
\label{deltaZ}\end{align}
where the contours $\Gamma^>_0,\Gamma^<_0$ encircle the origin clockwise with
$|z|>|\zeta|, |z|<|\zeta|$ respectively, and $\Gamma_\zeta$
encircles $\zeta$ but not the origin. 
Also 
\begin{align}
[Q^I_{0\hskip2pt J}, Y^K(\zeta)\} &= t^I_J[\zeta] Y^K(\zeta)
=\delta^K_J Y^I(\zeta) -{1\over 8} (-1)^{I+K} \delta^I_J
Y^K(\zeta) - {1\over 8} \delta^I_J Y^K(\zeta).
\label{deltaY}\end{align}
(\ref{deltaZ}),(\ref{deltaY}) correspond to infinitesimal field
transformations that leave the twistor string world sheet action invariant,
and are generated by the super commutator of the Noether charge with the field. 

Similarly, for any field $V(\zeta) \equiv V(Z(\zeta),Y(\zeta))$ 
that is a function of $Z_K(\zeta)$ and or $Y^K(\zeta)$,
such as the vertex operators, 
the operator product with the ordinary current is
\begin{align}
J^I_{0J}(z) V(\zeta)& \sim (z-\zeta)^{-1}\,
t^I_{\hskip2pt J}[\zeta] \hskip4pt V(\zeta)\; ,
\end{align}
From (\ref{tdef}),  it follows that
\begin{align}
[\; t^I_{\hskip2pt J}[\zeta], \;t^K_{\hskip2pt L}[\zeta]\;\}
&= f^{I\hskip3pt K\hskip14pt N}_{\hskip3pt J\hskip6pt L\,M}
\; t^M_{\hskip2pt N}[\zeta],
\nonumber\end{align}
where the structure constants 
$f^{I\hskip3pt K\hskip14pt N}_{\hskip3pt J\hskip6pt L\,M}$
are given (\ref{ssc}).
The charge acts 
\begin{align}
[Q^I_{0\hskip2pt J}, V(\zeta) \} = t^I_{\hskip2pt J}[\zeta] \; V(\zeta)\,,
\label{chargeV}\end{align}
and on a product of fields as 
\begin{align}
&[Q^I_{0\hskip2pt J}, V_1(\zeta_1) V_2(\zeta_2)\}\cr
&= [Q^I_{0\hskip2pt J}, V_1(Z(\zeta_1))\} \hskip5pt V_2(Z(\zeta_2)) 
+ (-1)^{(I+J)\text{deg}(V_1)}V_1(Z(\zeta_1)) \hskip5pt  
[Q^I_{0\hskip2pt J}, V_2(Z(\zeta_2))\} \cr
&= t^I_{\hskip2pt J}[\zeta_1] V_1(\zeta_1)\hskip5pt   V_2(\zeta_2)
+  (-1)^{(I+J)\text{deg}(V_1)} V_1(\zeta_1)\hskip5pt  t^I_{\hskip2pt J}
[\zeta_2] V_2(\zeta_2).
\end{align} 
This can be written as the standard Lie superalgebra comultiplication,
\begin{align}
\Delta Q^I_{0\hskip2pt J} &= 
Q^I_{0\hskip2pt J} \otimes 1 + (-1)^{(I+J)F}\otimes Q^I_{0\hskip2pt J},
\label{coprod}\end{align}
which instructs one how to build the action of the charge on two fields 
from knowledge of the action of the charge on a single field, etc. 
The operator $F$ is the fermion number operator.
Clearly the standard comultiplication follows from
the fact that the action of the charge 
on a single site is given by a commutator as in (\ref{chargeV}). 

\vskip20pt
\subsection{\it Action of 0rdinary Charges on the Tree Amplitudes}

It follows directly that the psl(4$|$4)
symmetry charges annihilate the tree amplitudes, as we now show.
The charges, which can be expressed in term of the modes as 
\begin{align}
Q^I_{0\hskip2pt J} = \sum_n Y^I_{-n} Z_{J\, n}
-{1\over 8} (-1)^{I+E}\delta^I_J \sum_n Y^E_{-n} Z_{E\, n} 
- {1\over 8}\delta^I_J \sum_n Y^E_{-n} Z_{E\, n},
\label{qandqdagger}\end{align} annihilate 
the vacuum,
\begin{align}
Q^I_{0\hskip2pt J} |0\rangle, \qquad \langle 0|
(Q^I_{0\hskip2pt J})^\dagger = \pm \langle 0| Q^I_{0\hskip2pt J}= 0,
\label{qannih}\end{align}
since the modes $Y^I_n, Z_n^I$ satisfy the vacuum conditions \cite{DG1}
\begin{align}
Y^I_n\vac=0, \quad n\geq 0, \qquad Z^I_n\vac=0,\quad n\geq 1,
\label{modes}\end{align}
where $Z(z) = \sum Z_n z^{-n}$, and $Y(z) = \sum_n Y_n z^{-n-1},$
and the hermiticity conditions are
\begin{align}\left(Z_{J\,n}\right)^\dagger &=Z_{J\,-n}, \qquad
(Y^I_n)^\dagger = - (-1)^I Y^I_{-n}.
\label{herm}\end{align}
This is in contrast to the $U(1)_R$ charge which does not 
annihilate the vacuum.
\vskip10pt

The tree amplitudes in twistor string theory are given by
\begin{align}
{\cal A}_n^{\rm tree} =\int \langle 0| e^{dq_0}V_1(z_1)
V_2(z_2)\ldots V_n (z_n) | 0\rangle 
\prod_{r=1}^ndz_r\Big/d\gamma_M 
d\gamma_S\label{tree}
\end{align}
where $d\gamma_M$ is the invariant measure on the M\"obius group, 
$d\gamma_S$ 
is the invariant measure on the group of scale transformations,
$d$ is the instanton number which is equal to 
one less than the number of negative helicities, 
$q_0 = \sum_{I=1}^8 q_0^I$ the sum of
the zero modes of $X^I$ relevant for bosonizing $Y^I(z), Z_J(z)$;
and the homogeneous conformal fields
$V_i(z_i)$ are the vertex operators of the gluon or graviton
supermultiplets \cite{BW, DG1, DI}.

Under a charge $Q_0$, 
the vertex operators transform as
\begin{align}
V(z)\rightarrow V'(z) &= e^{\rho Q_0} V(z) e^{-\rho Q_0}
\simeq V(z) + \rho [Q_0 , V(z)]  + \O(\rho^2).
\label{symtran2}\end{align} 

The psl(4$|$4) charges $Q_0$ commute with $e^{dq_0}$,  
\begin{align}
[Q_0, e^{dq_0}] = 0,
\label{Qq}\end{align}
since the modes satisfy
\begin{align}
Y^I_{n-d}e^{dq_0}=e^{dq_0}Y^I_n,\qquad Z^I_{n+d}e^{dq_0}=e^{dq_0}Z^I_n\qquad
\hbox{for } 1\leq I\leq 8. 
\end{align}
Therefore the tree amplitudes are invariant under the symmetry
transformations,
\begin{align}
{\cal A}_n^{\rm tree} &=\int \langle 0| e^{dq_0}V'_1(z_1)
V'_2 (z_2)\ldots V'_n (z_n) | 0\rangle
\prod_{r=1}^ndz_r\Big/d\gamma_M d\gamma_S\cr
&=\int \langle 0| e^{dq_0} e^{\rho Q_0}V_1(z_1) 
V_2 (z_2) \ldots V_n (z_n) | 0\rangle
\prod_{r=1}^ndz_r\Big/d\gamma_M d\gamma_S\cr
&=\int \langle 0| e^{dq_0} V_1(z_1)
V_2 (z_2)\ldots V_n (z_n) | 0\rangle
\prod_{r=1}^ndz_r\Big/d\gamma_M d\gamma_S
\label{invtree}\end{align}
from (\ref{qandqdagger}), (\ref{symtran2}), (\ref{Qq}).
Thus working to first order in $\rho$, we see that the
charges $Q_0$ annihilate the tree amplitudes, 
\begin{align}
\int \langle 0| e^{dq_0} Q_0 V_1(z_1)
V_2 (z_2)\ldots V_n (z_n) | 0\rangle
\prod_{r=1}^ndz_r\Big/d\gamma_M d\gamma_S = 0.
\label{chargeamp}\end{align}
Furthermore we can evaluate the action of the charges on the amplitude as 
follows. 

For simplicity, we consider the gluon tree amplitudes. 
The vertices for the  negative  and  positive helicity gluons are
\begin{align}
V^A_-(z) &=
\int  {dk}k^3\prod_{a=1}^2\delta(k\lambda^a(z)-\pi^a)
e^{ik\mu^{\dot a}(z)\bar\pi_{\dot a}}J^A(z)
\psi^1(z)\psi^2(z)\psi^3(z)\psi^4(z),\cr
V^A_+(z) &=
\int  {dk\over k}\prod_{a=1}^2\delta(k\lambda^a(z)-\pi^a)
e^{ik\mu^{\dot a} (z)\bar\pi_{\dot a}}J^A(z).
\end{align}
These are functions of $Z_J(z)$, whose
components are labeled as 
$Z^I = (\lambda^a,$ $\mu^{\dot a},$ $\psi^1,$
$\psi^2,\psi^3,\psi^4)$.

As an example, consider the action of the momentum operator on the 
vertices,
\begin{align}
[p_a^{\hskip3pt \dot a}, V^A_{\pm}(z)]
&= t^{\dot a}_{\hskip3pt a} [z]\hskip4pt  V^A_{\pm}(z)
= - \lambda_a(z) {\partial\over \partial \mu_{\dot a}(z)}
\hskip 4pt V^A_{\pm}(z) = 
 i k \bar\pi^{\dot a} \lambda_a(z) \hskip4pt  V^A_{\pm}(z)
\cr &= i \pi_a\bar\pi^{\dot a} \hskip4pt V^A_{\pm}(z).
\end{align}
Then from (\ref{chargeamp}), 
\begin{align}
&\int \langle 0| e^{dq_0} p_a^{\hskip3pt \dot a}  V_1(z_1)
V_2 (z_2)\ldots V_n (z_n) | 0\rangle
\prod_{r=1}^ndz_r\Big/d\gamma_M d\gamma_S\cr
&= \int \langle 0| e^{dq_0} V_1(z_1) \ldots
\sum_{r=1}^n [p_a^{\hskip3pt \dot a},  V_r(z_r)]\ldots
\ldots V_n (z_n) | 0\rangle
\prod_{r=1}^ndz_r\Big/d\gamma_M d\gamma_S\cr 
& =  i \left( \sum_{r=1}^n \pi_{ra}\bar\pi_r^{\dot a} \right) 
\hskip4pt  \int \langle 0| e^{dq_0} V_1(z_1)
V_2 (z_2)\ldots V_n (z_n) | 0\rangle
\prod_{r=1}^ndz_r\Big/d\gamma_M d\gamma_S\cr
&= 0,
\end{align}
which is momentum conservation. 

If we Fourier transform the gluon vertex operators
from $\bar \pi_{\dot a}$ back to $\omega^{\da}$ 
which is a twistor space coordinate, and introduce
the superspace coordinate $\theta^M$ to include all fields in 
the gluon supermultiplet, 
they become the wavefunction for the world sheet fields $Z^I(z)$ to be at
a point $Z'^I = (\pi^a,\omega^{\dot a}, \theta^M)$ in super twistor space, 
\begin{align}
W^A(z) &=\int \prod_{a=1}^2\delta(k\lambda^a(z)-\pi^a)
\delta(k\mu^{\dot a} (z)-\omega^{\dot a})\prod_{M=1}^4
(k\psi^M(z)-\theta^M){dk\over k} J^A(z).
\label{W}\end{align}
See for example \cite{DG1}.
Then the action of the charges on $W^A$ is
\begin{align}
[Q^I_{0\hskip3pt J}, W^A(z)]
&= t^I_{\hskip3pt J}[z]\hskip4pt  W^A(z) 
= {t'}^I_{\hskip3pt J}\hskip4pt  W^A(z) 
\end{align}
where ${t'}^I_{\hskip3pt J}$ 
is now expressed in terms of {\it points in twistor space}
rather than world sheet fields,
\begin{align} 
{t'}^{\hskip2pt I}_{\hskip2pt J} &\equiv
-(-1)^{IJ+I} \,  Z'_J {\partial\over\partial Z'_I}
+{1\over 8} \delta^I_J (-1)^{I+E} Z'_E{\partial\over
\partial Z'_E}
+{1\over 8} \delta^I_J Z'_E{\partial\over
\partial Z'_E}.
\label{tprime}\end{align}
So if we consider the 
superspace tree amplitude, that is the tree constructed with vertices
(\ref{W}), then the psl(4$|$4) charges act as 
\begin{align}
&\int \langle 0| e^{dq_0} Q^I_{0\hskip3pt J}  W_1(z_1)
W_2 (z_2)\ldots W_n (z_n) | 0\rangle
\prod_{r=1}^ndz_r\Big/d\gamma_M d\gamma_S\cr
&=\int \langle 0| e^{dq_0} W_1(z_1) \ldots
\sum_{r=1}^n [Q^I_{0\hskip3pt J},  W_r(z_r)]\ldots
\ldots W_n (z_n) | 0\rangle
\prod_{r=1}^ndz_r\Big/d\gamma_M d\gamma_S\cr
&= \sum_{r=1}^n  {t'}^I_{rJ}
\int \langle 0| e^{dq_0} W_1(z_1) W_2(z_2) \ldots W_n (z_n) | 0\rangle
\prod_{r=1}^ndz_r\Big/d\gamma_M d\gamma_S,\cr
&=0,\label{QW}\end{align}
which is zero because the charge 
annihilates the vacuum as in (\ref{chargeamp}).
A similar formula for the gauge field theory appears in \cite{Drummond2}.
Indeed the transformation on generators ${t'}^I_{rJ}$,
\begin{align}
- (-1)^{IJ+I} Z'_{rJ} { \partial\over\partial Z'_{rI}}
\rightarrow {Z'}_r^I { \partial\over\partial {Z'}_r^J}
\label{finalalg}\end{align}
(with trace and supertrace removed)
leaves the psl(4$|$4) commutation relations invariant,
and is the transformation of a matrix $M^I_J$,
$M\rightarrow - (M)^{\rm{supertranspose}}$ which is an outer automorphism
of the superalgebra psl(4$|$4).

Note that the tree amplitude $\M_n(\pi^a,\omega^{\da},\theta^M)$
is the Fourier transform of $M_n(\pi^a,\bpi_{\da}, \eta_{M})$, 
\begin{align}
\M_n(\pi^a,\omega^{\da},\theta^M) &\equiv
\int \langle 0| e^{dq_0}W^{A_1}(z_1)
W^{A_2}(z_2) \ldots W^{A_n}(z_n) | 0\rangle
\prod_{r=1}^n dz_r\Big/d\gamma_M d\gamma_S\cr
&=\int \prod_{r=1}^n \prod_{\da =1}^2\prod_{M=1}^4
d\eta_{rM} d\bpi_{r\da} e^{-i\omega_r^{\da}\bpi_{r\da}} e^{\theta^M_r\eta_{rM}}
M_n(\pi^a,\bpi_{\da}, \eta_{M}),
\end{align}
where $M_n(\pi^a,\bpi_{\da}, \eta_{M})$
is of Grassmann degree $4(d+1)$ and
is expressed in terms of some of the conjugate twistor space points  
$W'_I = (\bar \omega_a, \bpi_{\da}, \eta_{M})$, e.g.,
for MHV trees  
$M^{\rm MHV}_n(\pi^a,\bpi_{\da}, \eta_{M})
= \delta^4 (\sum_{r=1}^n\pi_r^a\bpi_{r\da}) 
\delta^8(\sum_{r=1}^n \pi_r^a\eta_{rM}){f^{A_1\ldots A_n}\over
\langle 12\rangle\ldots \langle n1\rangle}.$

\section{Non-local Noether currents and charges}
\label{sec:yangian}

In this section we compute non-local currents whose associated charges
will also annihilate the scattering amplitudes.
We show how these charges have
the same comultiplication rules as the level one generators of the 
Yangian superalgebra of psl(4$|$4).
In order to realize the coproduct of the Yangian, which is a
Hopf algebra, it will 
be necessary to define currents in terms of a field
that is not local with respect to the conformal fields \cite{Bernard2},
\cite{DB}.
We introduce the non-local field $\chi^I_{\hskip3pt J} (z)$,
\begin{align}
\chi^I_{\hskip3pt J} (z) &\equiv
\int^z_P dw J^I_{0\hskip2pt J}(w),
\label{chitilde}\end{align}
which not only is defined as an integral over a local field,
but also is not local with respect to local fields, {\it e.g.}
\begin{align}
(-1)^{R(I+J)}\; Z_R(\zeta) \; \chi^I_{\hskip3pt J} (z)
&\cong \chi^I_{\hskip3pt J} (z) \;Z_R(\zeta) 
\; + \; 2\pi i\,  [Q^I_{0\hskip2pt J}, Z_R(\zeta)\},
\label{loc}\end{align} 
where the left hand side of (\ref{loc}) is defined for 
$|\zeta|>|z|, |\zeta|>|P| $, the right hand side is defined for 
$|z|>|\zeta|, |P|>|\zeta|$ for an arbitrary point $P$,
and the equality $\cong$ is meant in the sense
of analytic continuation, see Figure 2. 
\begin{figure}[!h]
	\label{fig:fig1}
\centering%
\psfrag{P}{P}
\psfrag{w}{\fbox{$w$-plane}}
\psfrag{z}{z}
\psfrag{xi}{$\xi$}
\psfrag{c1}{$\mathcal{C}_{w_2}$}
\psfrag{c2}{$\mathcal{C}_{w_1}$}
\centering
\includegraphics[width=10cm]{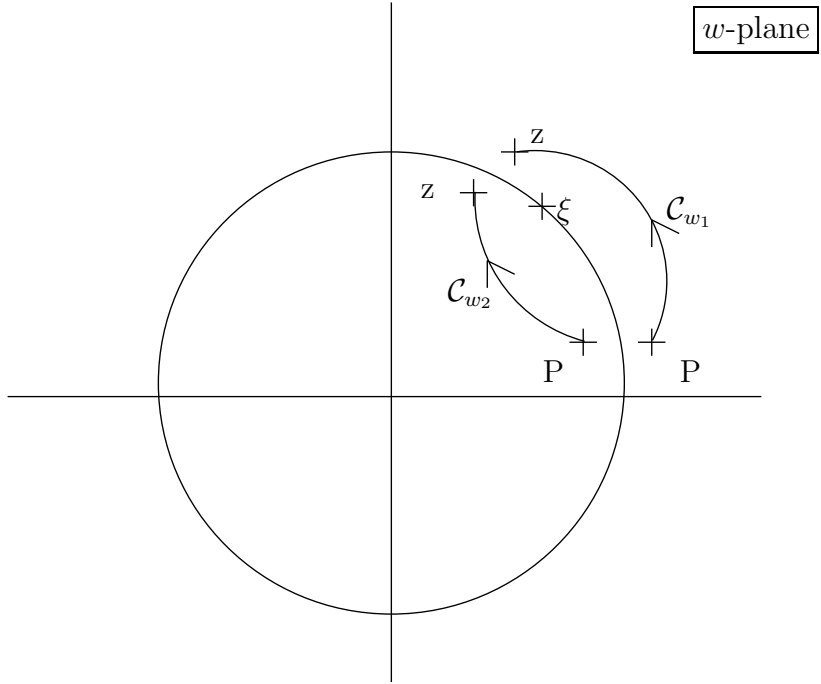}
\caption{\em The integral that defines
$\chi^I_{\ J}$ is along $\mathcal{C}_{w_1}$ for the right-hand side of \eqref{loc} and along
$\mathcal{C}_{w_2}$ for the left-hand side.}
\end{figure}

The operator product of  $\chi^I_{\hskip3pt J} (z)$
with $Z_R(\zeta)$ is
\begin{align}
\chi^I_{\hskip3pt J} (z) \; Z_R(\zeta)
\sim \int^z_P dw (w-\zeta)^{-1}\,
t^{\hskip2pt I}_{\hskip3pt J} [\zeta] \, Z_R(\zeta)
= \ln{(z-\zeta)\over (P-\zeta)}\;\;
t^{\hskip2pt I}_{\hskip3pt J} [\zeta] Z_R(\zeta).
\end{align}
We define the non-local currents 
$J^M_{1\hskip2pt N}(z)$ by
%from (\ref{NLJ}),
\begin{align}%\label{eq:levelonecurrent}
J^M_{1\hskip2pt N}(z) & =
f^{M\hskip7pt L\hskip6pt J}_{\hskip5pt
N K\hskip5pt I}  \int^z_P dw
J^I_{0\hskip2pt J}(w)\;J^K_{0\hskip2pt L}(z)
= f^{M\hskip7pt L\hskip6pt J}_{\hskip5pt
N K\hskip5pt I} \; \chi^I_{\hskip3pt J} (z)\;
J^K_{0\hskip2pt L}(z).
\label{levelonecurrent}\end{align}
It is not necessary to normal order because the singular terms
vanish due to the properties of the structure constants. 
The $J^M_{1N}(z)$ are not local with respect to the fields,

\begin{align}
(-1)^{R(M+N)}\; Z_R(\zeta) \;J^M_{1\hskip2pt N}(z) 
\cong &J^M_{1\hskip2pt N}(z)  \;
Z_R(\zeta)  + \cr
&+f^{M\hskip7pt L\hskip6pt J}_{\hskip5pt
N K\hskip5pt I}\; (-1)^{R(K+L)}\;
2\pi i \, [Q^I_{0\hskip2pt J}, Z_R(\zeta)\}\;
J^K_{0\hskip2pt L}(z). 
\end{align}
The operator product with fields $Z_R(\zeta)$ is
\begin{align}
J^M_{1\hskip2pt N}(z)  \; Z_R(\zeta)
&\sim (z-\zeta)^{-1} \; f^{M\hskip7pt L\hskip6pt J}_{\hskip5pt
N K\hskip5pt I} \; : \chi^I_{\hskip3pt J} (\zeta)
\; t^{\hskip2pt K}_{\hskip3pt L} [\zeta] \, Z_R(\zeta):\cr
&+  \ln{(z-\zeta)\over (P-\zeta)}\;\;
f^{M\hskip7pt L\hskip6pt J}_{\hskip5pt
N K\hskip5pt I} \; (-1)^{(I+J)(K+L)}\;
: J^K_{0\hskip2pt L}(z) \;\;
t^{\hskip2pt I}_{\hskip3pt J} [\zeta] \,Z_R(\zeta):.
\label{Yope}\end{align}
The action of the non-local charge on the field $Z_R(\zeta)$,
is
\begin{align}
Q^M_{1\hskip2pt N}\left(Z_R(\zeta)\right)
&= \oint_{\C_\zeta} dz 
J^M_{1\hskip2pt N}(z)  \; Z_R(\zeta)
 = 2 \; f^{M\hskip7pt L\hskip6pt J}_{\hskip5pt
N K\hskip5pt I} \; :\chi^I_{\hskip3pt J} (\zeta)
\;t^{\hskip2pt K}_{\hskip3pt L} [\zeta]\, Z_R(\zeta):,
\label{Yinf}\end{align}
where the cut for the logarithim extends from $\zeta$
passing through the point $P$, and the contour $\C_\zeta$ starts
just above $P$, circles around $\zeta$ and stops just below $P$.
See Figure\nobreak\,3. 
\begin{figure}[!h]
	\label{fig:fig2}
\centering%
\psfrag{P}{P}
\psfrag{C}{$\mathcal C_\zeta$}
\psfrag{z}{\fbox{$z$-plane}}
\psfrag{xi}{$\zeta$}
\centering
\includegraphics[width=10cm]{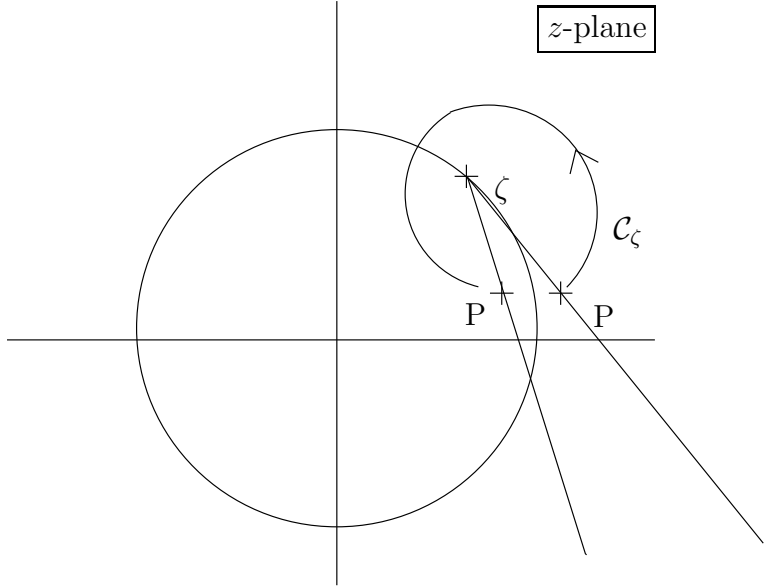}
\caption{\em The contour $\C_\zeta$ starts
just above $P$, circles around $\zeta$ and stops just below $P$.}
\end{figure}

To derive \eqref{Yinf} we have used
$\oint_{\C_\zeta} dz f(z) \ln{(z-\zeta)\over (P-\zeta)}\;\;
= -\int^\zeta_P dw f(w)$,
for any $f(w)$ analytic on and inside the contour $\C_\zeta$
\cite{Bernard2}.

The expression (\ref{Yinf}) is the infinitesimal transformation 
on the field $Z_R(\zeta)$
generated by the non-local charge \cite{DB}. It is not merely
$[Q^M_{1\hskip2pt N}, Z_R(\zeta) \}$ due to the lack of locality 
in (\ref{loc}),
\begin{align}
Q^M_{1\hskip2pt N}\left(Z_R(\zeta)\right)
= [ Q^M_{1\hskip2pt N}, Z_R(\zeta)\} 
 + f^{M\hskip7pt L\hskip6pt J}_{\hskip5pt
N K\hskip5pt I} \; (-1)^{R(K+L)}\;
2\pi i \,[ \widetilde Q^I_{0\hskip2pt J}, Z_R(\zeta)\} 
\; Q^K_{0\hskip2pt L}. 
\label{comYinf}\end{align}
This mismatch of the commutator and the infinitesimal field tranformation
follows from
\begin{align}
&Q^M_{1\hskip2pt N}\; Z_R(\zeta)
=\oint_{\C_1: |z|>|\zeta|,|P|>|\zeta|}dz \;
J^M_{1\hskip2pt N}(z) \; Z_R(\zeta),\cr
&(-1)^{R(M+N)}\; Z_R(\zeta) \; Q^M_{1\hskip2pt N}
= \oint_{\C_2: |\zeta|>|z|,|\zeta|>|P|}dz \;
(-1)^{R(M+N}\;Z_R(\zeta) \;
J^M_{1\hskip2pt N}(z)\cr
&= \oint_{\C_2: |\zeta|>|z|,|\zeta|>|P|}dz \;
J^M_{1\hskip2pt N}(z) Z_R(\zeta)
+ f^{M\hskip7pt L\hskip6pt J}_{\hskip5pt
N K\hskip5pt I} (-1)^{R(K+L)}\; 2\pi i \,
[Q^I_{0\hskip2pt J}, Z_R(\zeta)\}\;
Q^K_{0\hskip2pt L},
\end{align}
so that
\begin{align}
[ Q^M_{1\hskip2pt N}, Z_R(\zeta)\} &=
\oint_{\C_\zeta\equiv \C_1 - \C_2}dz
J^M_{1\hskip2pt N}(z) \; Z_R(\zeta) \cr
& 
- \ f^{M\hskip7pt L\hskip6pt J}_{\hskip5pt
N K\hskip5pt I} (-1)^{R(K+L)}\; 2\pi i \,[Q^I_{0\hskip2pt J}, Z_R(\zeta)
\}\;
Q^K_{0\hskip2pt L},
\end{align}
which is (\ref{comYinf}).

Therefore the operator product of the non-local current with 
any field $V(Z(\zeta))$ that is a function of $Z_R(\zeta),$
such as the gluon vertex operator, is
\begin{align}
J^M_{1\hskip2pt N}(z)  \; V(Z(\zeta))
& \sim (z-\zeta)^{-1} \; f^{M\hskip7pt L\hskip6pt J}_{\hskip5pt
N K\hskip5pt I} \; : \chi^I_{\hskip3pt J} (\zeta)
\; t^{\hskip2pt K}_{\hskip3pt L} [\zeta] \, V(Z(\zeta)):\cr
&+  \ln{(z-\zeta)\over (P-\zeta)}\;\;
f^{M\hskip7pt L\hskip6pt J}_{\hskip5pt
N K\hskip5pt I} \; (-1)^{(I+J)(K+L)}\;
: J^K_{0\hskip2pt L}(z) \;\;
t^{\hskip2pt I}_{\hskip3pt J} [Z(\zeta)] V(Z(\zeta)):.
\cr\label{YopeV}\end{align}
The action of the first level Yangian charge on the field $V(Z(\zeta))$,
is
\begin{align}
Q^M_{1\hskip2pt N}\left(V(Z(\zeta))\right)
&= \oint_{\C_\zeta} dz
J^M_{1\hskip2pt N}(z)  \; V(Z(\zeta))\cr
& = 2 \; \widetilde f^{M\hskip7pt L\hskip6pt J}_{\hskip5pt
N K\hskip5pt I} \; :\chi^I_{\hskip3pt J} (\zeta)
\;t^{\hskip2pt K}_{\hskip3pt L} [\zeta]\,  V(Z(\zeta)):
%&\qquad + {1\over 8} \big(-1)^M - (-1)^N\Big) : \chi^I_{\hskip3pt I} (\zeta)
%\;t^{\hskip2pt M}_{\hskip3pt N} [\zeta]\;  V(Z(\zeta)):. 
\label{YinfV}\end{align}

For simplicity, we consider a bosonic vertex operator $V(Z(\zeta))$,
where %Then from (\ref{loc}), (\ref{nonlocJ}), (\ref{comYinf}), (\ref{YinfV}),
\begin{align}
V(Z(\zeta)) \;\chi^I_{\hskip3pt J} (z)
&\cong \chi^I_{\hskip3pt J} (z) \;V(Z(\zeta)) 
\; + \; 2\pi i \,[Q^I_{0\hskip2pt J}, V(Z(\zeta))],
\label{locV1}\\
V(Z(\zeta)) \;J^M_{1\hskip2pt N}(z)
&\cong J^M_{1\hskip2pt N}(z)  \;
V(Z(\zeta))  + f^{M\hskip7pt L\hskip6pt J}_{\hskip5pt
N K\hskip5pt I}\;
2\pi i \,[Q^I_{0\hskip2pt J}, V((\zeta))]\;
J^K_{0\hskip2pt L}(z),
\label{locV2}\\
Q^M_{1\hskip2pt N}\left(V(Z(\zeta))\right)
&=[ Q^M_{1\hskip2pt N}, V(Z(\zeta))]
 + f^{M\hskip7pt L\hskip6pt J}_{\hskip5pt
N K\hskip5pt I} \; 2\pi i \,[ \widetilde Q^I_{0\hskip2pt J}, V(Z(\zeta))]
\; Q^K_{0\hskip2pt L}.
\label{locV}\end{align}

Since the infinitesimal transformation (\ref{Yinf}) is not a simple
(anti)commutator (\ref{comYinf}), the action of the non-local charge on a 
product of fields will not be as simple as for the ordinary generators 
(\ref{coprod}), and leads to the Hopf superalgebra coproduct.

{\it Comultiplication}

The action of the non-local charge on two fields is 
\begin{align}
Q^M_{1\hskip2pt N}\Big(V_1(Z(\zeta_1))\,
V_2(Z(\zeta_2))\Big) &=
\oint_{\C_{\zeta_1,\zeta_2}}dz\, J^M_{1\hskip2pt N}(z)\,
V_1(Z(\zeta_1))\, V_2(Z(\zeta_2)),
\label{yco}\end{align}
where the contour $\C_{\zeta_1,\zeta_2}$ starts at $P$ above both cuts
(cut from $\zeta_1$ to $P$, and cut 
from $\zeta_2$ to $P$), encircles both
$\zeta_1, \zeta_2$ and stops below both cuts at $P$, see Figure 4.
\begin{figure}[!h]
	\label{fig:fig3}
\centering%
\psfrag{P}{P}
\psfrag{xi1}{$\zeta_1$}
\psfrag{xi2}{$\zeta_2$}
\centering
\includegraphics[width=6cm]{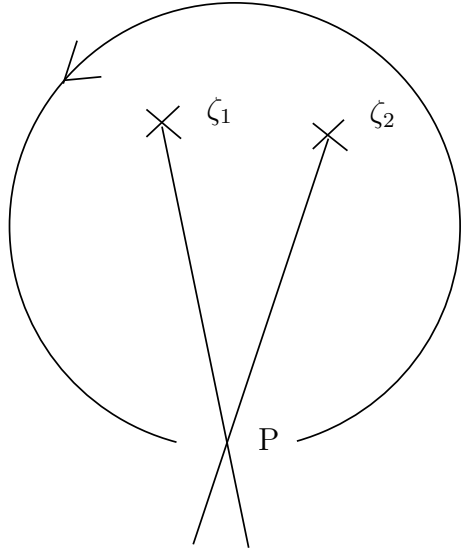}
\caption{\em The contour $\C_{\zeta_1,\zeta_2}$ starts at $P$ above both cuts
(cut from $\zeta_1$ to $P$, and cut 
from $\zeta_2$ to $P$), encircles both
$\zeta_1, \zeta_2$ and stops below both cuts at $P$.}
\end{figure}
For simplicity, we take both fields $V_1$ and $V_2$ to be bosonic.
We have for $|z|,|P| > |\zeta_1|,|\zeta_2|$
\begin{align}
J^M_{1\hskip2pt N}(z)\,V_1(Z(\zeta_1))\, V_2(Z(\zeta_2))
&= f^{M\hskip7pt L\hskip6pt J}_{\hskip5pt
N K\hskip5pt I}\;
\chi^I_{0\hskip2pt J}(z)\;
J^K_{0\hskip2pt L}(z)\; V_1(Z(\zeta_1))\, V_2(Z(\zeta_2))\, ,
\end{align}
and for $|\zeta_1|,|\zeta_2| > |z|,|P|$
\begin{align}
V_1(Z(\zeta_1))\, V_2(Z(\zeta_2))
\; J^M_{1\hskip2pt N}(z) &=
f^{M\hskip7pt L\hskip6pt J}_{\hskip5pt
N K\hskip5pt I}\;
\chi^I_{0\hskip2pt J}(z)\;
J^K_{0\hskip2pt L}(z)\; V_1(Z(\zeta_1))\, V_2(Z(\zeta_2)),\cr
&\hskip 5pt - f^{M\hskip7pt L\hskip6pt J}_{\hskip5pt
N K\hskip5pt I}\; 2\pi i \,[Q^I_{0\hskip2pt J}, V(Z(\zeta_1))]
\;V(Z(\zeta_2))\; J^K_{0\hskip2pt L}(z)\cr
&\hskip 5pt - f^{M\hskip7pt L\hskip6pt J}_{\hskip5pt
N K\hskip5pt I}\; V_1(Z(\zeta_1))\;
2\pi i \,[Q^I_{0\hskip2pt J}, V_2(Z(\zeta_2))] \;
J^K_{0\hskip2pt L}(z),
\label{ychargeop}\end{align}
using (\ref{locV1}) twice. 
Then
\begin{align}
[Q^M_{1\hskip2pt N},\,V_1(Z(\zeta_1))\, V_2(Z(\zeta_2))] &=
\oint_{\C_{\zeta_1,\zeta_2}}dz\, J^M_{1\hskip2pt N}(z)\,
V_1(Z(\zeta_1))\, V_2(Z(\zeta_2))\cr
&\hskip 5pt -  f^{M\hskip7pt L\hskip6pt J}_{\hskip5pt
N K\hskip5pt I}\; 2\pi i \, [Q^I_{0\hskip2pt J}, V_1(Z(\zeta_1))]
\;V_2(Z(\zeta_2))\; Q^K_{0\hskip2pt L}\cr
&\hskip 5pt - f^{M\hskip7pt L\hskip6pt J}_{\hskip5pt
N K\hskip5pt I}\; V_1(Z(\zeta_1))\;
2\pi i \,[Q^I_{0\hskip2pt J}, V_2(Z(\zeta_2))] \;
Q^K_{0\hskip2pt L},
\label{twofcom}\end{align}
So from (\ref{yco}),
\begin{align}
Q^M_{1\hskip2pt N}\Big(V_1(Z(\zeta_1))\,
V_2(Z(\zeta_2))\Big) &=
[Q^M_{1\hskip2pt N},\,V_1(Z(\zeta_1))\, V_2(Z(\zeta_2))]\cr
&\hskip5pt
+ f^{M\hskip7pt L\hskip6pt J}_{\hskip5pt
N K\hskip5pt I}\; 2\pi i \,[Q^I_{0\hskip2pt J}, V_1(Z(\zeta_1))]
\; Q^K_{0\hskip2pt L}\; V_2(Z(\zeta_2))\cr
&\hskip5pt
- f^{M\hskip7pt L\hskip6pt J}_{\hskip5pt
N K\hskip5pt I}\; 2\pi i \,[Q^I_{0\hskip2pt J}, V_1(Z(\zeta_1))]
\; [Q^K_{0\hskip2pt L}, V_2(Z(\zeta_2))]\cr
&\hskip 5pt + f^{M\hskip7pt L\hskip6pt J}_{\hskip5pt
N K\hskip5pt I}\; V_1(Z(\zeta_1))\;
2\pi i \,[Q^I_{0\hskip2pt J}, V_2(Z(\zeta_2))] \;
Q^K_{0\hskip2pt L}\cr
%&=
%\widetilde Q^M_{1\hskip2pt N}\Big(V_1(Z(\zeta_1))\Big)\; V_2(Z(\zeta_2))
%+ V_1(Z(\zeta_1))\; \widetilde Q^M_{1\hskip2pt N}\Big(V_2(Z(\zeta_2))\Big)\cr
%&\hskip 5pt - \widetilde f^{M\hskip7pt L\hskip6pt J}_{\hskip5pt
%N K\hskip5pt I}\;
%2\pi i \, [\widetilde Q^I_{0\hskip2pt J}, V_1(Z(\zeta_1))] \;
%[\widetilde Q^K_{0\hskip2pt L}, V_2(Z(\zeta_2))]\cr
&=
Q^M_{1\hskip2pt N}\Big(V_1(Z(\zeta_1))\Big)\; V_2(Z(\zeta_2))
+ V_1(Z(\zeta_1))\; Q^M_{1\hskip2pt N}\Big(V_2(Z(\zeta_2))\Big)\cr
&\hskip 5pt - f^{M\hskip7pt L\hskip6pt J}_{\hskip5pt
N K\hskip5pt I}\;
2\pi i \; Q^I_{0\hskip2pt J} \Big( V_1(Z(\zeta_1)) \Big)\;
Q^K_{0\hskip2pt L} \Big( V_2(Z(\zeta_2))\Big),
\label{twoftran}\end{align}
which can be written as the Hopf algebra comultiplication,
\begin{align}
\Delta Q^M_{1\hskip2pt N} &=
Q^M_{1\hskip2pt N} \otimes 1 + (-1)^{(M+N)F}\otimes 
Q^M_{1\hskip2pt N}
\hskip5pt - 2\pi i \; f^{M\hskip7pt L\hskip6pt J}_{\hskip5pt
N K\hskip5pt I}\; Q^I_{0\hskip2pt J} \otimes
Q^K_{0\hskip2pt L}.
\label{ycoprod}\end{align}
We can reexpress the action of the non-local charge on the product
of two fields as
\begin{align}
Q^M_{1\hskip2pt N}\Big(V(Z(\zeta_1))\,
V(Z(\zeta_2))\Big) &=
[Q^M_{1\hskip2pt N},  V(Z(\zeta_1))\,
V(Z(\zeta_2))]\cr
&\hskip10pt  + 
2\pi i f^{M\hskip7pt L\hskip6pt J}_{\hskip5pt
N K\hskip5pt I} [Q^I_{0\hskip2pt J},  V(Z(\zeta_1))\,
V(Z(\zeta_2))] \; Q^K_{0\hskip2pt L},
\label{Q1com}\end{align}
and will use this form 
to demonstrate how these non-local charges annihilate the tree amplitudes.

\section{More symmetry in the open twistor string}

We consider the action of the non-local charges on the 
tree amplitudes.
We show that the non-local charges  
annihilate the super gluon amplitudes of the open twistor string, and hence
give rise to interesting Ward identities
which we compare with $\mathcal N = 4$ Yang Mills field theory. 

The non-local charges formally annihilate the vacuum
using \eqref{modes},
\begin{align}
Q^M_{1\hskip2pt N} |0\rangle, \qquad \langle 0|
(Q^M_{1\hskip2pt N})^\dagger = \pm \langle 0| Q^M_{1\hskip2pt N},
\qquad [Q^M_{1N}, e^{dq_0}] = 0.
\label{q1annih}\end{align}
%\begin{align}
%[Q_1, e^{dq_0}] = 0.
%\label{Q1qtilde}\end{align}
In analogy with the ordinary $PSL(4|4)$ case, we consider the action of
the non-local charge on the tree amplitude
using (\ref{Q1com}), 
\begin{align}
&\int \langle 0| e^{dq_0} Q^M_{1\hskip3pt N}  \Big( W_1(z_1)
W_2 (z_2)\ldots W_n (z_n) \Big) | 0\rangle
\prod_{r=1}^ndz_r\Big/d\gamma_M d\gamma_S\label{Q1W1}\cr
&=\int \langle 0| e^{dq_0} [Q^I_{M\hskip3pt N},  W_1(z_1)
W_2 (z_2)\ldots W_n (z_n) ] | 0\rangle
\prod_{r=1}^ndz_r\Big/d\gamma_M d\gamma_S\cr
&\hskip10pt + 
\int \langle 0| e^{dq_0} 
f^{M\hskip3pt L\hskip14pt J}_{\hskip3pt N\hskip6pt K\,I}
\, 2\pi i \, [Q^I_{0\hskip3pt J},  W_1(z_1)
W_2 (z_2)\ldots W_n (z_n) ] \; Q^K_{0\hskip3pt L}| 0\rangle
\prod_{r=1}^ndz_r\Big/d\gamma_M d\gamma_S\cr
&=0,
\end{align}
where we have suppressed the group index.
To display the action of the non-local charge on the 
amplitudes, we compute (\ref{Q1W1}) directly, defining
\begin{align}
W(z_1,\ldots, z_n)\, \equiv \,  \langle 0| e^{dq_0}   \Big( W_1(Z(z_1))
\ldots  W_n(Z(z_n))\Big)  | 0\rangle.
\end{align}
We use (\ref{twoftran}) to express (\ref{Q1W1}),
\begin{align}
&\int \langle 0| e^{dq_0} Q^M_{1\hskip3pt N}  \Big( W_1(z_1)
W_2 (z_2) \ldots W_n(z_n)\Big)  | 0\rangle
\prod_{r=1}^ndz_r\Big/d\gamma_M d\gamma_S\cr
&= \int \langle 0| e^{dq_0} 
\sum_{j=1}^n \,W_1(z_1) \ldots Q^I_{1\hskip3pt J}
\Big(W_j (z_j)\Big)\; \ldots W_n(z_n)| 0\rangle
\prod_{r=1}^n dz_r\Big/d\gamma_M d\gamma_S\cr
&\hskip5pt - \int \langle 0| e^{dq_0} \;
2\pi i \, f^{M\hskip7pt L\hskip6pt J}_{\hskip5ptN K\hskip5pt I}\;
\sum_{1\le i<j\le n}\,  W_1(z_1) \ldots W_{i-1}(z_{i-1})
\; Q^I_{0\hskip2pt J} \Big( W_i(z_i)\Big) 
\, W_{i+1}(z_{i+1})\ldots \cr
&\hskip100pt \times W_{j-1}(z_{j-1})\,
Q^K_{0\hskip2pt L} \Big( W_j(z_j)\Big) W_{j+1}(z_{j+1}\; \ldots
W_n(z_n)| 0\rangle
\prod_{r=1}^ndz_r\Big/d\gamma_M d\gamma_S\cr
&=\, - 2 f^{M\hskip7pt L\hskip6pt J}_{\hskip5ptN K\hskip5pt I}\;
\sum_{1\le i<j\le n} t'^I_{i\,\hskip3pt J} \;
t'^K_{j\,\hskip3pt L}  
\; \int \; \Big( \ln{(z_i-z_j)\over (P-z_j)}
- \ln{(z_j-z_i)\over (P-z_i)}\Big) \;
W(z_1, \ldots, z_n)\prod_{r=1}^ndz_r\Big/d\gamma_M d\gamma_S\cr
%&\hskip5pt 
%+ 2 f^{M\hskip7pt L\hskip6pt J}_{\hskip5ptN K\hskip5pt I}\;
%\sum_{1\le i<j\le n} t'^I_{i\,\hskip3pt J} \;
%t'^K_{j\,\hskip3pt L}
%\; \int \; \ln{(z_j-z_i)\over (P-z_i)}\; 
%W(z_1,\ldots, z_n)\prod_{r=1}^ndz_r\Big/d\gamma_M d\gamma_S\cr
&\hskip5pt + ( 2\pi i)\,
f^{M\hskip7pt L\hskip6pt J}_{\hskip5ptN K\hskip5pt I}\;
\; \sum_{1\le i<j\le n}
t'^I_{i\,\hskip3pt J} \;t'^K_{j\,\hskip3pt L}
\; \int  W(z_1,\dots, z_n)
\prod_{r=1}^ndz_r\Big/d\gamma_M d\gamma_S\cr
% &\hskip5pt - 2\pi i\,
%f^{M\hskip7pt L\hskip6pt J}_{\hskip5ptN K\hskip5pt I}\;
%\; \sum_{1\le i<j\le n} t'^I_{i\,\hskip3pt J} \;
%t'^K_{j\,\hskip3pt L}  
%\; \int  W(z_1,\ldots, z_n)
%\prod_{r=1}^ndz_r\Big/d\gamma_M d\gamma_S
\label{acteval}\end{align}
where we have replaced the
$t^I_{\hskip 2pt i J}[z_i]$ which appear in 
evaluating the operator products, by $t'^I_{\hskip 2pt  iJ}$
defined in (\ref{tprime}), since the vertex operators $W(z)$ are delta functions
given in (\ref{W}).
To show the independence of $P$, we write \eqref{acteval} as 
\begin{align}
& - 2f^{M\hskip7pt L\hskip6pt J}_{\hskip5ptN K\hskip5pt I}\;
\sum_{1\le i<j\le n}
t'^I_{i\,\hskip3pt J} \;t'^K_{j\,\hskip3pt L}\; 
\int \; \Big( \ln{(z_i-z_j)\over (z_j-z_i)} 
- \ln{(P-z_j)\over (P-z_i)} \Big) \; W(z_1,\ldots, z_n)
\prod_{r=1}^ndz_r\Big/d\gamma_M d\gamma_S\cr
&\hskip5pt + 2\pi i\,
f^{M\hskip7pt L\hskip6pt J}_{\hskip5ptN K\hskip5pt I}\;
\; \sum_{1\le i<j\le n}
t'^I_{i\,\hskip3pt J} \;t'^K_{j\,\hskip3pt L}
\; \int  W(z_1,\ldots ,z_n)
\prod_{r=1}^ndz_r\Big/d\gamma_M d\gamma_S,
\label{q1af}\end{align}
where the $P$-dependent part is evaluated as\begin{align}
&f^{M\hskip7pt L\hskip6pt J}_{\hskip5ptN K\hskip5pt I}
\sum_{1\le i<j \le n} t'^I_{i\,\hskip3pt J} \;t'^K_{j\,\hskip3pt L}\;
\ln{(P-z_j)\over (P-z_i)} \, W(z_1,\ldots, z_n)\cr
%&=f^{M\hskip7pt L\hskip6pt J}_{\hskip5ptN K\hskip5pt I}
%\sum_{1\le i<j \le n} \Big(t'^I_{i\,\hskip3pt J} \; t'^K_{j\,\hskip3pt L}
%\;\ln(P-z_j) - t'^I_{i\,\hskip3pt J} \; t'^K_{j\,\hskip3pt L} \;
%\ln(P-z_j) \Big) \, W(z_1,\ldots, z_n)\cr
& = -\ln (P-z_1) \; f^{M\hskip7pt L\hskip6pt J}_{\hskip5ptN K\hskip5pt I}
\; \; t'^I_{1\,\hskip3pt J}
\sum_{j=2}^n t'^K_{j\,\hskip3pt L}
\, W(z_1,\ldots, z_n)\cr
& \hskip10pt
+ \sum_{\ell=2}^{n-1} \ln (P-z_{\ell}) \;
f^{M\hskip7pt L\hskip6pt J}_{\hskip5ptN K\hskip5pt I}
\Big( \sum_{i=1}^{\ell -1} t'^I_{i\,\hskip3pt J} \;
t'^K_{\ell\,\hskip3pt L} - t'^I_{\ell\,\hskip3pt J}
\; \sum_{j=\ell +1}^n t'^K_{j\,\hskip3pt L}\Big)\;
\, W(z_1,\ldots, z_n)\cr
& \hskip10pt +\ln (P-z_n) \;
f^{M\hskip7pt L\hskip6pt J}_{\hskip5ptN K\hskip5pt I}
\sum_{i=1}^{n-1} t'^I_{i\,\hskip3pt J} \;
t'^K_{n\,\hskip3pt L}
\, W(z_1,\ldots, z_n).
\label{PP}\end{align}
We use that the ordinary symmetry annihilates
the integrand of the amplitude (\ref{QW}), and
\begin{align}
f^{M\hskip7pt L\hskip6pt J}_{\hskip5ptN K\hskip5pt I}\;
\Big(t'^I_{i\,\hskip3pt J} \;t'^K_{j\,\hskip3pt L} + t'^I_{j\,\hskip3pt J}
\;t'^K_{i\,\hskip3pt L}\Big)=0 \quad \hbox{for $i\ne j$},
\end{align}
and for each $i$,
\begin{align}f^{M\hskip7pt L\hskip6pt J}_{\hskip5ptN K\hskip5pt I}\;
\Big(t'^I_{i\,\hskip3pt J} \;t'^K_{i\,\hskip3pt L}\Big)&=
{1\over 2} f^{M\hskip7pt L\hskip6pt J}_{\hskip5ptN K\hskip5pt I}\;
f^{I\hskip3pt K\hskip14pt S}_{\hskip3pt J\hskip6pt L\,R}\;
t'^R_{i\,\hskip3pt S} =0,\end{align}
since the Killing form (\ref{vkf}) vanishes.
Then \eqref{PP} becomes
\begin{align}
& = \ln (P-z_1) \; f^{M\hskip7pt L\hskip6pt J}_{\hskip5ptN K\hskip5pt I}
\; \; t'^I_{1\,\hskip3pt J}
 t'^K_{1\,\hskip3pt L}
\, W(z_1,\ldots, z_n)\cr
& \hskip10pt
+ \sum_{\ell=2}^{n-1} \ln (P-z_{\ell}) \;
f^{M\hskip7pt L\hskip6pt J}_{\hskip5ptN K\hskip5pt I}
\Big( \sum_{i=1}^{\ell -1} t'^I_{i\,\hskip3pt J} \;
t'^K_{\ell\,\hskip3pt L}
+ t'^I_{\ell\,\hskip3pt J} \;
t'^K_{\ell\,\hskip3pt L}
+ t'^I_{\ell\,\hskip3pt J}
\; \sum_{j=1}^{\ell -1} t'^K_{j\,\hskip3pt L}\Big)\;
\, W(z_1,\ldots, z_n)\cr
& \hskip10pt +\ln (P-z_n) \;
f^{M\hskip7pt L\hskip6pt J}_{\hskip5ptN K\hskip5pt I}
\sum_{i=1}^{n-1} t'^I_{i\,\hskip3pt J} \;
t'^K_{n\,\hskip3pt L}
\, W(z_1,\ldots, z_n)\cr
&=0,\end{align}
Therefore the $P$ dependent part of (\ref{q1af}) vanishes,
and using $\ln{(z_i-z_j)\over (z_j-z_i)}\; = \ln(-1) = \pi i + 2\pi i p$
for any integer $p$, we find 
(\ref{acteval}) leads to
\begin{align}
&\int \langle 0| e^{dq_0} Q^I_{1\hskip3pt J}
\Big( W_1(z_1)\ldots W_n(z_n)\Big)  | 0\rangle
\prod_{r=1}^ndz_r\Big/d\gamma_M d\gamma_S\cr
&= -4\pi i p \; f^{M\hskip7pt L\hskip6pt J}_{\hskip5ptN
K\hskip5pt I}\;\sum_{1\le i<j\le n}
t'^I_{i\,\hskip3pt J} \;t'^K_{j\,\hskip3pt L}\;
\int \langle 0| e^{dq_0} \;W_1(Z(z_1)) \ldots\; W_n(Z(z_n)\, | 0\rangle
\prod_{r=1}^ndz_r\Big/d\gamma_M d\gamma_S\cr&=0,
\end{align}
for arbitrary integer $p$, so along with the ordinary 
symmetries (\ref{QW}), we have
\begin{align}f^{M\hskip7pt L\hskip6pt J}_{\hskip5ptN K\hskip5pt I}
\sum_{1\le i<j \le n} t'^I_{i\,\hskip3pt J} \;t'^K_{j\,\hskip3pt L}\;
\int \langle 0| e^{dq_0} \;W_1(Z(z_1)) \ldots  W_n(Z(z_n)\,
| 0\rangle\prod_{r=1}^ndz_r\Big/d\gamma_M d\gamma_S=0,\label{ysymts}
\end{align}
which reflects the non-local symmetry of the tree amplitudes of the
twistor string. 
We note that up to central terms that separately annihilate
the amplitudes, 
a similar formula 
appears in \cite{Drummond2}-\cite{DF2}
as the first level Yangian generators acting on gauge theory amplitudes.

\section{PSL(4$|$4) Yangian superalgebra symmetry}
\label{sec:yangian}

In this section, we show how the open twistor string carries a 
realization of the Yangian of psl(4$|$4). 
This realization will be independent of $P$, and  
is inspired from the observation that the Ward identities
for the non-local charges are independent of $P$, \eqref{ysymts}.

Let us start by computing the OPE of our non-local currents 
\eqref{levelonecurrent} 
with the Virasoro field \eqref{VA},
\begin{align}\label{eq:levelonecurrentvirasoro}
L_{YZ}(z) J^M_{1\hskip2pt N}(\zeta) & = \frac{J^M_{1\hskip2pt N}(\zeta)}
{(z-\zeta)^2}
+ \frac{\del J^M_{1\hskip2pt N}(\zeta)}{(z-\zeta)}+
f^{M\hskip7pt L\hskip6pt J}_{\hskip5ptN K\hskip5pt I} 
\frac{ J^I_{0\hskip2pt J}(P)\;J^K_{0\hskip2pt L}(w)}{(z-P)}\, .
\end{align}
Hence, the non-local current is a Virasoro primary only in the 
limit $P\rightarrow\infty$. 
Note that taking this limit does not commute with the integrals we take. 
But we take it as a hint for a P-indendent representation of the Yangian.

We keep the action of the level zero charges as before and choose the level 
one generators to act trivially on a single field
\begin{equation}
Q_{1\ N}^M\big(V(z)\big) \ = \ 0\, ,
\label{onef}\end{equation}
while the action of $Q_{1\ N}^M$ 
on products of fields is defined by the Yangian comultiplication 
\eqref{ycoprod}, so that on a product of two fields,
\begin{align}
Q^M_{1\hskip2pt N}\Big (V(z_1) V(z_2) \Big)&=
- 2\pi i \; f^{M\hskip7pt L\hskip6pt J}_{\hskip5pt
N K\hskip5pt I}\; Q^I_{0\hskip2pt J}\Big(V(z_1)\Big) \otimes
Q^K_{0\hskip2pt L}\Big(V(z_2)\Big).
\label{yycoprod}\end{align}
The crucial observation is that this representation of the Yangian 
also provides a symmetry of the super gluon amplitudes of the 
open twistor string, as can be seen by \eqref{ysymts}.

We prove that the tree level representation in terms of twistor string fields 
for the PSL(4$|$4) charges \eqref{noech}, \eqref{noecu}, \eqref{coprod}, 
and the level one charges, \eqref{onef}, \eqref{yycoprod}, \eqref{ycoprod},
is consistent with the Serre relation,
because it satisfies a useful criterion \cite{DNW2} as we now show.

\vskip10pt

{\it Serre relation and the useful criterion}

The Serre relation for the Yangian of $psl(4|4)$ can be written as
\begin{align}
& f^{M\hskip3pt R\hskip14pt L}_{\hskip3pt N\hskip6pt S\,K}
[Q^I_{1\hskip2pt J},\, Q^K_{1\hskip2pt L}\}
+ (-1)^{(I+J)(M+N+R+S)}
\, f^{R\hskip3pt I\hskip14pt L}_{\hskip3pt S\hskip6pt J\,K}
[Q^M_{1\hskip2pt N},\,Q^K_{1\hskip2pt L}\}
+\cr
&+(-1)^{(R+S)(I+J+M+N)}
\,f^{I\hskip3pt M\hskip14pt L}_{\hskip3pt J\hskip6pt N\,K}
[Q^R_{1\hskip2pt S},\, Q^K_{1\hskip2pt L}\}\, =\cr
&\qquad\qquad\quad\qquad= h (-1)^{(C+D)(K+L) + (G+H)(U+V) + G + H}\;\times\cr
&\qquad\qquad\qquad\quad\qquad\times\,
f^{\,I\hskip6pt D\hskip6pt Q}_{\hskip5pt J C\hskip5pt P}\;
f^{\,M\hskip6pt F\hskip6pt L}_{\hskip5pt N E\hskip5pt K}\;
f^{R\hskip3pt G\hskip14pt V}_{\hskip3pt S\hskip6pt H\,U}\;
f^{C\hskip3pt E\hskip14pt H}_{\hskip3pt D\hskip6pt F\,G}\;
{\bf\Big \{} Q^P_{0\hskip2pt Q}\;Q^K_{0\hskip2pt L}
\;Q^U_{0\hskip2pt V}{\bf \Big ]},
\label{widehatserre}\end{align}
where ${\bf\big \{}\ldots{\bf \big ]}$ is the graded totally symmetrized
product and $h$ is a constant that depends on the normalization.

It is sufficient to check the Serre relation acting on one field,
since then the coproduct \eqref{ycoprod}
assures the relation for all higher sites. Clearly the
left-hand side of \eqref{widehatserre} vanishes for one field from
\eqref{onef}.

We show in appendix \ref{app:rhs} that the right-hand side of the Serre 
relation acting on one field also vanishes, using the fact that 
our representation for the $gl(4|4)$ charge
$\widetilde Q^I_{0\hskip2pt J}$ on a single field,
\begin{align}
[\widetilde Q^I_{0\hskip2pt J}, V(Z(\zeta))] &=
[\oint dz :Y^I(z) Z_J(z): , V(Z(\zeta))]\cr & =
\Big(-(-1)^{IJ+I} \, \left( Z_J(\zeta) {\partial\over
\partial Z_I(\zeta)}\right)   +  Y^I(\zeta) {\partial\over\partial Y^J(\zeta)}
\Big) \;V(Z(\zeta)) 
\equiv 
\widetilde t^I_{\hskip2pt J}[\zeta] \; V(Z(\zeta)),
\cr\end{align}
satisfies a useful criterion \cite{DNW2}:
\begin{align}
(-1)^E \widetilde Q^M_{0\hskip2pt E}\,\widetilde Q^E_{0\hskip2pt J} 
\big( V(Z(\zeta)\big)
= \widetilde Q^M_{0\hskip2pt J}  \big( V(Z(\zeta)\big).
\label{criter1}\end{align}
Here we consider $V(Z(z)$ as a homogeneous function
%so that
%\begin{align}
$Z_E(z) {\partial\over \partial Z_E(z)} V(Z(z)) = 0$
%\end{align}
which describes conformal fields $V(z)$ such as the
vertex operators $W(z)$. 
This will simplify our discussion, although is not necessary since the
central terms we drop would cancel in what follows.
To derive (\ref{criter1}),
\begin{align}
&(-1)^E \widetilde Q^M_{0\hskip2pt E}\, \widetilde Q^E_{0\hskip2pt J} 
\big( V(Z(\zeta)\big)
= (-1)^E \widetilde t^M_{0\hskip2pt E}\,\widetilde t^E_{0\hskip2pt J} \, 
V(Z(\zeta)\cr
&= (-1)^{ME+M+EJ} Z_E(\zeta) {\partial\over \partial Z_M(\zeta)}
\, \big ( Z_J(\zeta) {\partial\over \partial Z_E(\zeta)} \,
V(Z(\zeta)) \Big) \cr
&= (-1)^{M+EJ+MJ} Z_E(\zeta) Z_J(\zeta)\,
{\partial\over \partial Z_E(\zeta)} {\partial\over \partial Z_M(\zeta)}\,
V(Z(\zeta) 
= -(-1)^{M+MJ} Z_J(\zeta)\,
{\partial\over \partial Z_M(\zeta)}\,V(Z(\zeta) \cr
&= \widetilde Q^M_{0\hskip2pt J}  \big( V(Z(\zeta)\big).
\label{dercriter}\end{align}

From \eqref{onef} and \eqref{ycoprod} we also have that
\begin{align}
[Q^I_{0\hskip2pt J},
Q^K_{1\hskip2pt L}\}
&= f^{I\hskip3pt K\hskip14pt S}_{\hskip3pt J\hskip6pt L\,R}
\, Q^R_{1\hskip2pt S},
\label{defrelone}\end{align}
acting on gauge invariant states.
Therefore with the proof of \eqref{widehatserre} in appendix E, 
the Yangian symmetry algebra now follows from the defining relations
\eqref{chainv},  \eqref{defrelone} and \eqref{widehatserre}.

%%%%%%%%%%%%%%%%%%%%%%%%%%%%%%%%%%%%%%%%%%%%%%%%%%%%%%%%%%%%%%%%%%%%%
\subsection*{\bf Acknowledgments}

We are grateful to Nima Arkani-Hamed, Jacob Bourjaily,  
Freddy Cachazo, James Drummond, Peter Goddard, Johaness Henn, 
Hubert Saleur, David Skinner and Edward Witten for discussions.
This work was partially supported by the U.S. Department of Energy,
Grant No. DE-FG02-06ER-4141801, Task A.

\appendix

\section{Some super analysis}\label{app:super}

We provide some properties of Lie superalgebras and Lie supergroups 
we use in our analysis.
A good reference for Lie supergroups is \cite{Berezin:1987wh}, 
in this appendix we use the notation of \cite{Creutzig:2009zz}.

\subsection{\emph{The Lie superalgebra psl(N$|$N)}}\label{app:psl44}

The Lie superalgebra sl(N$|$N) can be represented by matrices
$(T)_{IJ} =  \left(\begin{array}{cc}M_1&M_2\\
M_3&M_4 \end{array}\right)$, with \Str \, (T) $\equiv$ Tr$M_1$ - Tr $M_4$ $=0$. 
This superalgebra possesses a one-dimensional ideal generated by the identity matrix.
The quotient of sl(N$|$N) by this ideal is the Lie superalgebra psl(N$|$N).
It is convenient to work with the Lie superalgebra sl(N$|$N), we simply have to remember that
we divided out the central element.

We define the psl(4$|$4) structure constants by
\begin{equation}
[T^a, T^b\} = f^{ab}_{\hskip6pt c}\,  T^c = 
f^{abc} g_{cd} T^d\,, \label{scomrel}
\end{equation}
where the brackets denote either commutators or anticommutators.
There is an invariant,
nondegenerate metric $g_{ab}$  that is used to raise and lower indices,
$g^{ab}=\half \Str \,T^aT^b$ and $g_{ab}=\half \Str \,T^bT^a$.  
The structure constants $f_{abc}$ are totally antisymmetric with
an additional minus sign under the interchange of two odd nearest 
neighbor indices. 

It can be useful to rewrite the single index superalgebra generators $J^a$ 
with a double index as $(E_{AB})_{IJ} = \delta_{AI}\delta_{BJ}$,
where 
$$[E_{AB}, E_{CD}] = \delta_{CB} E_{AD} - \delta_{AD} E_{CB},$$
$$[E_{AB}, E_{CD}\} = \delta_{CB} E_{AD} - (-1)^{\hbox{ (deg $E_{AB}$)
(deg $E_{CD}$)}} \delta_{AD} E_{CB},$$
where deg $E_{AB}$ is 0 for bosonic generators, and 1 for fermionic 
generators.

\subsection{\emph{Lie supergroups}}\label{app:supergroup}

The Lie supergroup is a Lie group over a Grassmann ring and it is obtained 
by exponentiating the Grassmannn envelope of the Lie superalgebra. An element 
of the Grassmann envelope has the following form
\begin{equation}
\rho \ = \ \rho_a T^a
\end{equation}
where the $\rho_a$ are Grassmann-even when $T^a$ is bosonic and Grassmann-odd 
when $T^a$ is fermionic.
Hermitian conjugation is complex conjugation of the transpose
in the bosonic case. For matrices generating a Lie superalgebra, it is super
complex conjugation of the supertranspose,
$$\rho^\ddagger  =  \bar\rho_a (\overline T^a)^{st},$$
where the bar is ordinary complex conjugation for bosons, and 
satisfies\footnote{Lifting complex conjugation to the ring of Grassmann numbers is not unique.
Our choice ensures that the adjoint operation for Lie supergroups is involutive.} 
$$ \overline{c\theta}  = \bar{c}\bar\theta, \qquad
\bar{\bar{\theta}} = -\theta, \qquad \overline{\theta_1\theta_2} 
= \bar\theta_1\bar\theta_2, 
$$
for any Grassmann elements $\theta, \theta_i$ and any complex number $c$.
The supertranspose of $T$  is defined 
\be\label{eq:supertranspose}
\begin{split}
\left(\begin{array}{cc}M_1 & M_2 \\ M_3 & M_4\\ \end{array}\right)^{st}
=\left(\begin{array}{cc}M^t & -M_3^t\\ M_2^t & M_4^t
\\\end{array}\right)
\end{split}
\nonumber\ee
so that $(TS)^{st} = S^{st} T^{st}$.
  
\section{Structure constants and their properties}
\label{app:structure}

The psl(4$|$4) structure constants are given by
\begin{align}
f^{I\hskip3pt K\hskip14pt N}_{\hskip3pt J\hskip6pt L\,M}
&= \delta^K_J\delta^N_L\delta^I_M - (-1)^{(I+J)(K+L)}
\, \delta^N_J\delta^I_L\delta^K_M
+ {1\over 8}  ((-1)^L-(-1)^K)
(-1)^I \delta^I_J\delta^N_L\delta^K_M
\cr &\hskip10pt
- {1\over 8}((-1)^J - (-1)^I) (-1)^K\, \delta^N_J\delta^K_L\delta^I_M
\hskip10pt - {1\over 8} \Big( 1 - (-1)^{I+J}\Big) \,
\delta^I_L\, \delta^K_J\, \delta^N_M.
\label{ssc}\end{align}
These have vanishing Killing form
\begin{align}
f^{M\hskip6pt L\hskip6pt J}_{\hskip5ptN K\hskip5pt I}
\;\; f^{I\hskip3pt K\hskip14pt S}_{\hskip3pt J\hskip6pt L\,R}
&= 0,
\label{vkf}\end{align}
and are traceless and supertraceless in all indices,
\begin{align}
\delta^J_I \;\; f^{I\hskip3pt K\hskip12pt N}_{\hskip3pt J\hskip6pt
L\;M} &= 0,\quad
(-1)^J \delta^J_I \;\;
f^{I\hskip3pt K\hskip12pt N}_{\hskip3pt J\hskip6pt L\;M}= 0,
\quad f^{I\hskip3pt K\hskip12pt N}_{\hskip3pt J\hskip6pt L\;M}
\;\;\;\delta^M_N = 0, \quad
f^{I\hskip3pt K\hskip12pt N}_{\hskip3pt J\hskip6pt L\;M}
\;\;\; (-1)^M \delta^M_N = 0,
\label{finalscprop}\end{align}
as well as totally anti supersymmetric in any pair of indices,
{\it for eg.} \begin{align}
f^{I\hskip3pt K\hskip12pt N}_{\hskip3pt J\hskip6pt
L\;M} = -(-1)^{(I+J)(K+L)} f^{K\hskip3pt I\hskip12pt N}_{\hskip3pt L
\hskip6pt J\;M}.
\nonumber\end{align}
They satisfy the Jacobi identity
\begin{align}
f^{I\hskip3pt K\hskip14pt S}_{\hskip3pt J\hskip6pt L\,R}
\,f^{R\hskip3pt W\hskip14pt V}_{\hskip3pt S\hskip6pt X\,U}
+ f^{K\hskip3pt W\hskip14pt S}_{\hskip3pt L\hskip6pt X\,R}
\,f^{R\hskip3pt I\hskip14pt V}_{\hskip3pt S\hskip6pt J\,U}
\,(-1)^{(I+J)(R+S)}
+ f^{W\hskip3pt I\hskip14pt S}_{\hskip3pt X\hskip6pt J\,R}
\,f^{R\hskip3pt K\hskip14pt V}_{\hskip3pt S\hskip6pt L\,U}
\,(-1)^{(W+X)(I+J+K+L)} =0.
\label{jac}\end{align}

The inverted structure constants are
\begin{align}
f^{M\hskip6pt L\hskip6pt J}_{\hskip5ptN K\hskip5pt I} &=
(-1)^K \left( \delta_N^L\delta_I^M\delta_K^J
-(-1)^{(M+N)(M+I)} \delta_K^M\delta_I^L\delta_N^J\right)\cr
&\hskip5pt
+{1\over 8} ( 1 - (-1)^{I+J})(-1)^M \delta^M_N\delta^L_I\delta^J_K
+{1\over 8} ( (-1)^I - (-1)^J) \delta^J_N\delta^L_K\delta^M_I\cr
&\hskip5pt + {1\over 8} \Big( (-1)^N - (-1)^M\Big) \,
\delta^M_K\delta^L_N\delta^J_I.
\label{isctilde}\end{align}

We `raise' and `lower' the index pairs on the structure constants by
\begin{align}
g^{A\hskip7pt C}_{\hskip4pt B\hskip7pt D} = (-1)^A \delta^A_D \, \delta^C_B,
\qquad (g^{-1})^{\hskip6pt B\hskip7pt D}_{A\hskip7pt C} 
= (-1)^B \delta^D_A \, \delta^B_C,
\end{align}
\begin{align}
f^{M\hskip6pt L\hskip6pt J}_{\hskip5ptN K\hskip5pt I}
&= f^{M\hskip3pt P\hskip12pt J}_{\hskip3pt N\hskip6pt Q\;I}
\; (g^{-1})^{\hskip6pt Q\hskip7pt L}_{P\hskip7pt K}
= f^{M\hskip3pt P\hskip12pt J}_{\hskip3pt N\hskip6pt Q\;I}
\; (-1)^Q \delta^Q_K \, \delta^L_P, \cr
f^{M\hskip3pt P\hskip12pt J}_{\hskip3pt N\hskip6pt Q\;I}
&= f^{M\hskip6pt L\hskip6pt J}_{\hskip5ptN K\hskip5pt I}
\; \;g^{K\hskip7pt P}_{\hskip4pt L\hskip7pt Q}
=  f^{M\hskip6pt L\hskip6pt J}_{\hskip5ptN K\hskip5pt I}
\; \; (-1)^K \delta^P_L \, \delta^K_Q. 
\end{align}
\section{Real form psl(4$|$4)}{\label{app:realform}

The Noether charge generators of the real form $psl(4|4)$ can be written as
\begin{align} p_a^{\hskip3pt \dot a} = Q^{\dot a}_{0\hskip2pt a} & =\,
- \oint dz  \; Y^{\dot a} (z) Z_a(z), \cr
k^a_{\hskip3pt \dot a} = Q^a_{0\hskip2pt \dot a}& =
- \oint dz  \; Y^a (z) Z_{\dot a} (z),\cr
 m^a_{\hskip3pt b} = Q^{a}_{0\hskip2pt b} &=
- \oint dz\; {1\over 2} (Y_b(z) Z^a(z) + Y^a(z) Z_b(z))\cr
&= - \oint dz \left( Y^a(z) Z_b(z) - \half\delta^a_b
Y^c(z) Z_c(z)\right) \cr
\widetilde m^{\da}_{\hskip3pt \db} = Q^{\da}_{0\hskip2pt \db} & =
- \oint dz\; {1\over 2}
(Y_{\db}(z) Z^{\da}(z) + Y^{\da}(z) Z_{\db}(z))\cr
&= - \oint dz \left( Y^{\dot a}(z) Z_{\dot b}(z)
- \half\delta^{\da}_{\db}
Y^{\dot c}(z) Z_{\dot c}(z)\right),\cr
d = Q_0^{(d)} &= - \oint dz\; {1\over 2}
(Y^{a}(z) Z_{a}(z) - Y^{\da}(z) Z_{\da}(z)), \cr
r^A_{\hskip3pt B} = Q^A_{0\hskip 3pt B}
&= -\oint dz\;   (Y^A(z) Z_B(z) - \quarter \delta^A_B \,Y^C(z) Z_C(z)),  \cr
q_a^{\hskip 3pt A} = Q^A_{0\hskip2pt a} &=
\oint dz  \; Y^A (z) Z_a (z),\cr
\widetilde q^{\da}_{\hskip 3pt A} = Q^{\da}_{0\hskip2pt A} &=
- \oint dz  \; Y^{\da} (z) Z_A (z),\cr
s^a_{\hskip 3pt A} = Q^a_{0\hskip2pt A} &=
- \oint dz  \; Y^a (z) Z_A (z),\cr
\widetilde s^A_{\hskip 3pt \da} = Q^A_{0\hskip2pt \da} &=
\oint dz  \; Y^A (z) Z_{\da} (z),\qquad\qquad\qquad
\label{charges}\end{align}
where $I = a,\da, A$, so 
$1\le a,\da\le 2$ and $1\le A\le 4$. All the generators in
(\ref{charges}) are antihermitian, except for $q_a^{\hskip3pt A}$ and
$\widetilde s^A_{\hskip5pt\dot a}$, which are hermitian.

The commutation relations for the generators (\ref{charges})
satisfy the real form $psl(4|4)$,  
\begin{align} [m^a_{\hskip3pt b}, J_c] & = \delta^a_c J_b -\half \delta^a_b J_c,
\quad\qquad [m^a_{\hskip3pt b}, J^c] = - \delta^c_b J^a
+\half \delta^a_b J^c,\cr
[\widetilde m^{\da}_{\hskip3pt \db}, J_{\dot c}] & = \delta^{\da}_{\dot c}
J_{\db} - \half \delta^{\da}_{\db} J_{\dot c},
\quad
\qquad [\widetilde m^{\da}_{\hskip3pt \db}, J^{\dot c}] =
-\delta^{\dot c}_{\db} J^{\da}
+\half\delta^{\dot a}_{\db} J^{\dot c},\cr
[r^A_{\hskip3pt B}, J_C] & = \delta^A_C J_B -\quarter\delta^A_B J_C,
\; \qquad [r^A_{\hskip3pt B}, J^C] = -\delta^C_B J^A
+ \quarter\delta^A_B J^C,\cr
[s^a_{\hskip3pt A}, p_b^{\hskip4pt \db}]
&= \delta^a_b \widetilde q^{\db}_{\hskip5pt A},\qquad
[k^a_{\hskip 4pt\da}, \widetilde q^{\db}_{\hskip4pt C}]
= -\delta_{\da}^{\db} s^a_{\hskip 3pt C} \;\Rightarrow \;
\hskip4pt [k^{a\da}, \widetilde q_{\db C}]
= \delta^{\da}_{\db} s^a_{\hskip 3pt C},\cr
[\widetilde s^A_{\hskip3pt \da}, p_b^{\hskip4pt \db}]
&= - \delta^{\db}_{\da} q^{A}_{\hskip5pt b} \;\Rightarrow
\; [\widetilde s^{A\da}, p_{b\db}]
=  \delta^{\da}_{\db} q^{A}_{\hskip5pt b},
\qquad\quad [k^a_{\hskip4pt \da}, q_b^{\hskip4pt  B}]
= \delta^a_b \widetilde s^B_{\hskip3pt \da},\cr
\{\widetilde q^{\da}_{\hskip3pt A}, q^B_{\hskip3pt b}&\}
= \delta^B_A p_b^{\hskip3pt \da}, \hskip140pt
\{\widetilde s^A_{\hskip3pt \da}, s^b_{\hskip3pt B}\}
= \delta^A_B k^b_{\hskip3pt \da},\cr
[k^a_{\hskip3pt \da}, p_{\dot b}^{\hskip3pt \db}]
&= -\delta^{\db}_{\da} m^a_{\hskip3pt b} +
\delta^a_b m^{\db}_{\hskip3pt \da} - \delta^{\db}_{\da} \delta^a_b  \, d\;
\; \Rightarrow\;\;
[k^{a\da}, p_{b\db}]
= \delta^{\da}_{\db} m^a_{\hskip3pt b} +
\delta^a_b \widetilde
m^{\da}_{\hskip3pt \db} + \delta^{\db}_{\da} \delta^a_b  \, d,\cr
\{ s^a_{\hskip3pt A}, q_b^{\hskip 4pt B}& \} =
\delta^B_A m^a_{\hskip3pt b} + \delta^a_b r^B_A
+\half  \delta^a_b \delta^B_A ( d - {\cal C}),\cr
\{\widetilde s^A_{\hskip3pt \da}, \widetilde q^{\db}_{\hskip3pt B}&\}
= \delta^A_B \widetilde m^{\db}_{\hskip 3pt \da}
- \delta^{\db}_{\da} r^A_{\hskip3pt B}
+ \half \delta^A_B \delta^{\db}_{\da} ( d + {\cal C}),\cr
& [d, J] = \hbox{dim} J, \qquad  [{\cal C}, J] = 0,
\label{psl}\end{align}
where the central generator
${\cal C} \equiv \half \oint dz Y^I(z)Z_I(z)$ acts as zero
on gauge invariant states.
%The non-zero conformal dimensions are
%$\hbox{dim}\; p_a^{\hskip3pt \da} = 1,
%\hbox{dim}\; k^a_{\hskip3pt \da} = -1,
%\hbox{dim}\; q_a^{\hskip3pt A} = \half,
%\break\hbox{dim}\; \widetilde q^{\da}_{\hskip3pt A} = \half,
%\hbox{dim}\; s^a_{\hskip3pt A} = -\half,
%\hbox{dim}\; \widetilde s^A_{\hskip4pt \da} = -\half.$

We raise and lower the bosonic indices as
$Z^a = \epsilon^{ab} Z_b,\;  Z_a = \epsilon_{ab} Z^b$, and
$Z^{\da} = \epsilon^{\da\db} Z_{\db}$, $Z_{\da} 
= \epsilon_{\da\db} Z^{\db}, $ with $\epsilon^{12} = 1 = -\epsilon^{21},\; 
\epsilon_{12} = -1 = -\epsilon_{21}$
for both the dotted and undotted indices.
 
\section{Vanishing of the right-hand side of the Serre relation
\eqref{widehatserre} on one field} \label{app:rhs}

We evaluate the right hand side of the Serre relation 
first for $gl(4|4)$,
\begin{align}
&(-1)^{(C+D)(K+L) + (G+H)(U+V) + G + H}\;
\widetilde f^{\,I\hskip6pt D\hskip6pt Q}_{\hskip5pt J C\hskip5pt P}\;
\widetilde f^{\,M\hskip6pt F\hskip6pt L}_{\hskip5pt N E\hskip5pt K}\;
\widetilde f^{R\hskip3pt G\hskip14pt V}_{\hskip3pt S\hskip6pt H\,U}\;
\widetilde f^{C\hskip3pt E\hskip14pt H}_{\hskip3pt D\hskip6pt F\,G}\;
{\bf\Big \{} \widetilde Q^P_{0\hskip2pt Q}\;\widetilde  Q^K_{0\hskip2pt L}
\;\widetilde  Q^U_{0\hskip2pt V}{\bf \Big ]}\cr
&= \bigg( - (-1)^{MJ+RN + S(M+J+R+N) + S + E}\,
\delta^I_S \,\Big\{  \widetilde Q^M_{0\hskip2pt E}\,
\widetilde Q^E_{0\hskip2pt J}\,
\widetilde Q^R_{0\hskip2pt N} - \widetilde Q^M_{0\hskip2pt J}\,
\widetilde Q^R_{0\hskip2pt E}\,
\widetilde Q^E_{0\hskip2pt N}\Big]\cr
&\hskip12pt +  (-1)^{(M +J)(N+R) + N + E}\;
%{RJ+MS + N(R+J+M+S) + N + E + (M+N)(R+S)}\,
\delta^I_N \,\Big\{  \widetilde Q^R_{0\hskip2pt E}\,
\widetilde Q^E_{0\hskip2pt J}\,
\widetilde Q^M_{0\hskip2pt S} - \widetilde Q^R_{0\hskip2pt J}\,
\widetilde Q^M_{0\hskip2pt E}\,
\widetilde Q^E_{0\hskip2pt S}\Big]\cr
&\hskip12pt +  (-1)^{(J+S)(R+N) + S + E} 
%{IN + RJ + S(I + N + R + J ) + S+E + (I+J) (M+N)}\,
\;\delta^M_S \,\Big\{  \widetilde Q^I_{0\hskip2pt E}\,
\widetilde Q^E_{0\hskip2pt N}\,
\widetilde Q^R_{0\hskip2pt J} - \widetilde Q^I_{0\hskip2pt N}\,
\widetilde Q^R_{0\hskip2pt E}\,
\widetilde Q^E_{0\hskip2pt J}\Big]\cr
&\hskip12pt -  (-1)^{JM + MS + JS + E} 
\;\delta^R_N \,\Big\{  \widetilde Q^I_{0\hskip2pt E}\,
\widetilde Q^E_{0\hskip2pt S}\,
\widetilde Q^M_{0\hskip2pt J} - \widetilde Q^I_{0\hskip2pt S}\,
\widetilde Q^M_{0\hskip2pt E}\,
\widetilde Q^E_{0\hskip2pt J}\Big]\cr
&\hskip12pt -  (-1)^{IR+RN+NI+E}\;
%{(I+N)(J+R)+J+E+(M+N)(I+J)}\;
\delta^M_J \,\Big\{  \widetilde Q^R_{0\hskip2pt E}\,
\widetilde Q^E_{0\hskip2pt N}\,
\widetilde Q^I_{0\hskip2pt S} - \widetilde Q^R_{0\hskip2pt N}\,
\widetilde Q^I_{0\hskip2pt E}\,
\widetilde Q^E_{0\hskip2pt S}\Big]\cr
&\hskip12pt +  (-1)^{IS+SN+NJ+JM+MI + E}\;
\delta^R_J \,\Big\{  \widetilde Q^M_{0\hskip2pt E}\,
\widetilde Q^E_{0\hskip2pt S}\,
\widetilde Q^I_{0\hskip2pt N } - \widetilde Q^M_{0\hskip2pt S}\,
\widetilde Q^I_{0\hskip2pt E}\,
\widetilde Q^E_{0\hskip2pt N}\Big]\Bigg),
\label{rss}\end{align}
which vanishes for our single site representation from
(\ref{criter1}).
This holds as well for $gl(n|n)$.
Here the gl(4$|$4) structure constants are those of psl(4$|$4) but
without the traces and supertraces removed,
\begin{align}
\widetilde f^{I\hskip3pt K\hskip14pt N}_{\hskip3pt J\hskip6pt L\,M}
&= \delta^K_J\delta^N_L\delta^I_M - (-1)^{(I+J)(K+L)}
\, \delta^N_J\delta^I_L\delta^K_M,\cr
\widetilde f^{M\hskip6pt L\hskip6pt J}_{\hskip5ptN K\hskip5pt I} &=
(-1)^K \left( \delta_N^L\delta_I^M\delta_K^J
-(-1)^{(M+N)(M+I)} \delta_K^M\delta_I^L\delta_N^J\right).
\end{align}
\eqref{rss} implies that that the psl(4$|$4) expression  
\begin{align}
(-1)^{(C+D)(K+L) + (G+H)(U+V) + G + H}\;
f^{\,I\hskip6pt D\hskip6pt Q}_{\hskip5pt J C\hskip5pt P}\;
f^{\,M\hskip6pt F\hskip6pt L}_{\hskip5pt N E\hskip5pt K}\;
f^{R\hskip3pt G\hskip14pt V}_{\hskip3pt S\hskip6pt H\,U}\;
f^{C\hskip3pt E\hskip14pt H}_{\hskip3pt D\hskip6pt F\,G}\;
{\bf\Big \{} Q^P_{0\hskip2pt Q}\;Q^K_{0\hskip2pt L}
\;Q^U_{0\hskip2pt V}{\bf \Big ]}
\label{rsshat}\end{align}
also vanishes on one site, for if we view (\ref{rss}) as
a tensor  ${\cal M}^{IMR}_{JNS}$, then
up to terms containing the
central generator $\sum_C \widetilde Q^C_{0C}$, (\ref{rsshat}) is just 
the tensor ${\cal M}^{IMR}_{JNS}$ with the 
trace and supertrace removed in each pair of indices $IJ,MN,RS$.

This follows from the properties of the psl(n$|$n)
structure constants given in appendix B,
\begin{align}
&(-1)^{(C+D)(K+L) + (G+H)(U+V) + G + H}\;
f^{\,I\hskip6pt D\hskip6pt Q}_{\hskip5pt J C\hskip5pt P}\;
f^{\,M\hskip6pt F\hskip6pt L}_{\hskip5pt N E\hskip5pt K}\;
f^{R\hskip3pt G\hskip14pt V}_{\hskip3pt S\hskip6pt H\,U}\;
f^{C\hskip3pt E\hskip14pt H}_{\hskip3pt D\hskip6pt F\,G}\;
{\bf\Big \{} Q^P_{0\hskip2pt Q}\;Q^K_{0\hskip2pt L}
\;Q^U_{0\hskip2pt V}{\bf \Big ]}\cr
&= (-1)^{(C+D)(K+L) + (G+H)(U+V) + G + H}\;
f^{\,I\hskip6pt D\hskip6pt Q}_{\hskip5pt J C\hskip5pt P}\;
f^{\,M\hskip6pt F\hskip6pt L}_{\hskip5pt N E\hskip5pt K}\;
f^{R\hskip3pt G\hskip14pt V}_{\hskip3pt S\hskip6pt H\,U}\;
\widetilde f^{C\hskip3pt E\hskip14pt H}_{\hskip3pt D\hskip6pt F\,G}\;
{\bf\Big \{} \widetilde Q^P_{0\hskip2pt Q}\; \widetilde Q^K_{0\hskip2pt L}
\;\widetilde Q^U_{0\hskip2pt V}{\bf \Big ]}\cr
&\sim (\widetilde f + h_1 + h_3) 
(\widetilde f + h_1 + h_3) (\widetilde f + h_1 + h_3)
\, \widetilde f\, \widetilde Q\,\widetilde Q\, \widetilde Q,
\label{hatserre}\end{align}
where we have given separate labels to the supertrace and trace terms in 
\eqref{ssc} and \eqref{isctilde}  for the three pair of indices 
as $h_1,  h_2,  h_3$,
$f = \widetilde f + h_1 + h_2 + h_3$, and
the expression in the final term of (\ref{hatserre}), 
$ (\widetilde f + h_1) (\widetilde f + h_1) (\widetilde f + h_1) \, 
\widetilde f\, \widetilde Q\, \widetilde Q\, \widetilde Q$ is
the tensor ${\cal M}^{IMR}_{JNS}$ with the
trace and supertrace removed in each pair of indices $IJ,MN,RS$.

Thus we have shown that the right hand side of 
the Serre relation (\ref{widehatserre}) 
vanishes for charges acting on a single field,
which together with \eqref{onef} and the coproduct \eqref{ycoprod} implies
that the Serre relation holds for any product of gauge invariant fields.

%%% References %%%

\providecommand{\bysame}{\leavevmode\hbox to3em{\hrulefill}\thinspace}
\providecommand{\MR}{\relax\ifhmode\unskip\space\fi MR }
\providecommand{\MRhref}[2]{%
  \href{http://www.ams.org/mathscinet-getitem?mr=#1}{#2}
}
\providecommand{\href}[2]{#2}


\begin{thebibliography}{99}
\bibitem{W}
E.~Witten,
{\it Perturbative gauge theory as a string theory in twistor space},
Commun.\ Math.\ Phys.\  {\bf 252}, 189 (2004)
[arXiv:hep-th/0312171].

\bibitem{B}
N.~Berkovits,
{\it An alternative string theory in twistor space for N = 4 super-Yang-Mills},
Phys.\ Rev.\ Lett.\  {\bf 93}, 011601 (2004)
[arXiv:hep-th/0402045].

\bibitem{BW} 
N.~Berkovits and E.~Witten,
{\it Conformal supergravity in twistor-string theory},
JHEP {\bf 0408}, 009 (2004)
[arXiv:hep-th/0406051].

\bibitem{DG3}
 L.~Dolan and P.~Goddard,
{\it Gluon tree amplitudes in open twistor string theory,}
JHEP {\bf 0912}, 032 (2009)
[arXiv:0909.0499 [hep-th]].

\bibitem{DG4}
L.~Dolan and P.~Goddard,
{\it General split helicity gluon tree amplitudes in open twistor string
theory,}
JHEP {\bf 1005}, 044 (2010)
[arXiv:1002.4852 [hep-th]].

\bibitem{Spradlin:2009qr}
M.~Spradlin and A.~Volovich,
{\it From twistor string theory to recursion relations},
Phys.\ Rev.\  D {\bf 80}, 085022 (2009)
[arXiv:0909.0229 [hep-th]].

\bibitem{Bourjaily:2010kw}
J.~L.~Bourjaily, J.~Trnka, A.~Volovich and C.~Wen,
{\it The Grassmannian and the twistor string: connecting all trees in N=4 SYM},
arXiv:1006.1899 [hep-th].

\bibitem{Nandan:2009cc}
D.~Nandan, A.~Volovich and C.~Wen,
{\it A Grassmannian etude in NMHV minors,}
JHEP {\bf 1007}, 061 (2010)
[arXiv:0912.3705 [hep-th]].

\bibitem{RMV1}
R.~Roiban, M.~Spradlin and A.~Volovich,
{\it Dissolving N = 4 loop amplitudes into QCD tree amplitudes},
Phys.\ Rev.\ Lett.\  {\bf 94}, 102002 (2005)
[arXiv:hep-th/0412265].

\bibitem{RMV2}
R.~Roiban, M.~Spradlin and A.~Volovich,
{\it A googly amplitude from the B-model in twistor space},
JHEP {\bf 0404}, 012 (2004)
[arXiv:hep-th/0402016].

\bibitem{RV}
R.~Roiban and A.~Volovich,
{\it All googly amplitudes from the B-model in twistor space},
Phys.\ Rev.\ Lett.\  {\bf 93}, 131602 (2004)
[arXiv:hep-th/0402121].

\bibitem{RMV3}
R.~Roiban, M.~Spradlin and A.~Volovich,
{\it On the tree-level S-matrix of Yang-Mills theory},
Phys.\ Rev.\  D {\bf 70}, 026009 (2004)
[arXiv:hep-th/0403190].

\bibitem{ACC}
N.~Arkani-Hamed, F.~Cachazo and C.~Cheung,
{\it The Grassmannian origin of dual superconformal invariance,}
JHEP {\bf 1003}, 036 (2010)
[arXiv:0909.0483 [hep-th]].

\bibitem{ACCK1}
N.~Arkani-Hamed, F.~Cachazo, C.~Cheung and J.~Kaplan,
{\it The S-Matrix in twistor space,}
JHEP {\bf 1003}, 110 (2010)
[arXiv:0903.2110 [hep-th]].

\bibitem{ACCK2}
N.~Arkani-Hamed, F.~Cachazo, C.~Cheung and J.~Kaplan,
{\it A Duality for the S Matrix},
JHEP {\bf 1003}, 020 (2010)
[arXiv:0907.5418 [hep-th]].

\bibitem{ACBT1}
N.~Arkani-Hamed, J.~Bourjaily, F.~Cachazo and J.~Trnka,
{\it Local spacetime physics from the Grassmannian,}
arXiv:0912.3249 [hep-th].

\bibitem{ABCT2}
N.~Arkani-Hamed, J.~Bourjaily, F.~Cachazo and J.~Trnka,
{\it Unification of residues and Grassmannian dualities,}
arXiv:0912.4912 [hep-th].

\bibitem{Mason}
L.~Mason and D.~Skinner,
{\it Dual superconformal invariance, momentum twistors and grassmannians},
JHEP {\bf 0911}, 045 (2009),
[arXiv:0909.0250 [hep-th]].

\bibitem{MasonandSkinner}
L.~Mason and D.~Skinner,
{\it Heterotic twistor-string theory},
Nucl.\ Phys.\  B {\bf 795} (2008) 105
[arXiv:0708.2276 [hep-th]].

\bibitem{Skinner}
D.~Skinner,
{\it A direct proof of BCFW recursion for twistor-strings},
arXiv:1007.0195 [hep-th].

\bibitem{Brand}
A.~Brandhuber, P.~Heslop and G.~Travaglini,
{\it A note on dual superconformal symmetry of the N=4 super Yang-Mills
S-matrix},
Phys.\ Rev.\  D {\bf 78}, 125005 (2008)
[arXiv:0807.4097 [hep-th]].

\bibitem{Wolf}
M.~Wolf,
{\it On hidden symmetries of a super gauge theory and twistor string theory,}
JHEP {\bf 0502}, 018 (2005)
[arXiv:hep-th/0412163].

\bibitem{DNW1}{L. Dolan, C. Nappi, and E. Witten,
{\it A relation between approaches to integrability in superconformal
Yang-Mills theory}, JHEP{\bf 0310}, 017 (2003)
[arXiv:hep-th/0308089].}

\bibitem{DNW2}{L. Dolan, C. Nappi, and E. Witten,
{\it Yangian symmetry in D = 4 superconformal Yang-Mills theory},
Contributed to 3rd International Symposium on Quantum Theory and Symmetries 
(QTS3), Cincinnati, Ohio, 10-14 Sep 2003.
Published in *Cincinnati 2003, Quantum theory and symmetries* 300-315,
[arXiv:hep-th/0401243].} 

\bibitem{MZ}
J.~A.~Minahan and K.~Zarembo,
{\it The Bethe-ansatz for N = 4 super Yang-Mills,}
JHEP {\bf 0303}, 013 (2003)
[arXiv:hep-th/0212208].

\bibitem{Bena}{I. Bena, J. Polchinski and R. Roiban, {\it Hidden
symmetries of the AdS(5) x $S^5$ superstring,}
Phys.\ Rev.\  D {\bf 69}, 046002 (2004)
[arXiv:hep-th/0305116].}

\bibitem{NMtwo}
N.~J.~MacKay,
{\it Introduction to Yangian symmetry in integrable field theory,}
Int.\ J.\ Mod.\ Phys.\  A {\bf 20}, 7189 (2005)
[arXiv:hep-th/0409183].

\bibitem{NM}{N. J. MacKay, {\it On the classical origins of Yangian symmetry
in integrable field theory,} Phys. Lett. {\bf B 281} 90 (1992);
Erratum-ibid.\ B {\bf 308}, 444 (1993).}

\bibitem{DRone}{V. Drinfel'd, {\it Hopf algebras and the quantum Yang-Baxter
equation}, Sov. Math. Dokl. {\bf 32} 254 (1985).}

\bibitem{DRtwo}{V. Drinfel'd, {\it A new realization of Yangians and
quantized affine algebras},  Sov. Math. Dokl. {\bf 36} 212 (1988).}

\bibitem{Dolan:2003yv}
{L.~Dolan,
{\it Yangians and Kac-Moody loop algebras in superconformal gauge theory,}
Proceedings of Science, 
27th Johns Hopkins Workshop on Current Problems in Particle Theory: 
Symmetries and Mysteries of M-Theory, Goteborg, Sweden, 24-26 Aug 2003,
http://pos.sissa.it/cgi-bin/reader/conf.cgi?confid=11}


\bibitem{D1}{L. Dolan,
{\it Kac-Moody algebra is hidden symmetry of chiral models},
\ Phys.\ Rev. \ Lett. {\bf 47} 1371 (1981).}

\bibitem{DR}{L. Dolan and A. Roos,
{\it Nonlocal currents as Noether currents},
Phys.\ Rev.\  D {\bf 22}, 2018 (1980).}

\bibitem{D2} {L.~Dolan,
{\it Kac-Moody algebras and exact solvability in hadronic physics},
Phys.\ Rept.\  {\bf 109}, 1 (1984).}

\bibitem{Brezin}
E.~Brezin, C.~Itzykson, J.~Zinn-Justin and J.~B.~Zuber,
{\it Remarks about the existence of non-local charges in two-dimensional
models,} Phys.\ Lett.\  B {\bf 82} (1979) 442.

\bibitem{Polyakovtwo}{A. M. Polyakov,
{\it Gauge fields as rings of glue,} Nucl. Phys. {\bf B164}, 1971 (1980).}

\bibitem{LP}{M. Luscher and K. Pohlmeyer, {\it Scattering of massless
lumps and nonlocal charges in the two-dimensional classical
non-linear sigma model},  Nucl. Phys. {\bf B137}, 46 (1978).}

\bibitem{LP2}{
M. Luscher, {\it Quantum non-local charges and absence of particle
production in the two-dimensional non-linear $\sigma$ model},
\ Nucl.\  Phys. {\bf B135}, 1 (1978).}

\bibitem{pop}{A.~D.~Popov and C.~R.~Preitschopf,
{\it Conformal symmetries of the self-dual Yang-Mills equations},
Phys.\ Lett.\ B {\bf 374}, 71 (1996)
[arXiv:hep-th/9512130].}

\bibitem{Drummond1}
J.~M.~Drummond, J.~Henn, G.~P.~Korchemsky and E.~Sokatchev,
{\it Dual superconformal symmetry of scattering amplitudes in N=4
super-Yang-Mills theory},
Nucl.\ Phys.\  B {\bf 828}, 317 (2010)
[arXiv:0807.1095 [hep-th]].

\bibitem{DHSS2}
J.~M.~Drummond, J.~Henn, V.~A.~Smirnov and E.~Sokatchev,
{\it Magic identities for conformal four-point integrals,}
JHEP {\bf 0701}, 064 (2007)
[arXiv:hep-th/0607160].

\bibitem{AM}
L.~F.~Alday and J.~M.~Maldacena,
{\it Gluon scattering amplitudes at strong coupling},
JHEP {\bf 0706}, 064 (2007)
[arXiv:0705.0303 [hep-th]].

\bibitem{Berkovits:2008ic}
N.~Berkovits and J.~Maldacena,
{\it Fermionic T-Duality, dual superconformal symmetry, and the
amplitude/Wilson loop connection},
JHEP {\bf 0809}, 062 (2008)
[arXiv:0807.3196 [hep-th]].

\bibitem{Beisert:2008iq}
N.~Beisert, R.~Ricci, A.~A.~Tseytlin and M.~Wolf,
{\it Dual superconformal symmetry from AdS5 x S5 superstring integrability,}
Phys.\ Rev.\  D {\bf 78}, 126004 (2008)
[arXiv:0807.3228 [hep-th]].

\bibitem{NB}
N.~Beisert,
{\it On Yangian symmetry in planar N=4 SYM},
arXiv:1004.5423 [hep-th].

\bibitem{Hodges:2009hk}
A. Hodges,
{\it Eliminating spurious poles from gauge-theoretic amplitudes,}
arXiv:0905.1473 [hep-th].

\bibitem{Drummond2}{
J.~M.~Drummond, J.~M.~Henn and J.~Plefka,
{\it Yangian symmetry of scattering amplitudes in N=4 super Yang-Mills theory},
JHEP {\bf 0905}, 046 (2009)
[arXiv:0902.2987 [hep-th]].}

\bibitem{DF}{
J.~M.~Drummond and L. Ferro,
{\it Yangians, Grassmannians and T-duality},
JHEP {\bf 1007}, 027 (2010)
[arXiv:1001.3348 [hep-th]].}

\bibitem{DF2}{
J.~M.~Drummond and L. Ferro,
{\it The Yangian origin of the Grassmannian integral},
[arXiv:1002.4622 [hep-th]].}

\bibitem{Bernard2}{
D.~2.~Bernard, Z.~Maassarani and P.~Mathieu,
{\it Logarithmic Yangians in WZW models},
Mod.\ Phys.\ Lett.\  A {\bf 12}, 535 (1997)
[arXiv:hep-th/9612217].}

\bibitem{DB}{D. Bernard, {\it Hidden Yangians in 2D massive current algebras,}
Comm. Math. Phys. {\bf 137}, 191 (1991).}

\bibitem{DG1} {L. Dolan, and P. Goddard,
{\it Tree and loop amplitudes in open twistor string theory},
JHEP {\bf 0706}, 005 (2007) [arXiv:hep-th/0703054].}

\bibitem{DG2}
L.~Dolan and P.~Goddard,
{\it Current algebra on the torus}, Commun.
Math. Phys. {\bf 285}, 219 (2009)
[arXiv:0710.3743 [hep-th]].

\bibitem{Creutzig:2009zz}
 T.~Creutzig,
{\it Branes in supergroups,}
arXiv:0908.1816 [hep-th].

\bibitem{Creutzig:2010zp}
T.~Creutzig and Y.~Hikida,
{\it Branes in the OSP(1$|$2) WZNW model},
arXiv:1004.1977 [hep-th].

\bibitem{Creutzig:2008ag}
 T.~Creutzig,
{\it Geometry of branes on supergroups,}
Nucl.\ Phys.\  B {\bf 812}, 301 (2009)
[arXiv:0809.0468 [hep-th]].

\bibitem{Creutzig:2008ek}
T.~Creutzig and V.~Schomerus,
{\it Boundary correlators in supergroup WZNW models,}
Nucl.\ Phys.\  B {\bf 807}, 471 (2009)
[arXiv:0804.3469 [hep-th]].

\bibitem{Creutzig:2007jy}
T.~Creutzig, T.~Quella and V.~Schomerus,
{\it Branes in the GL(1$|$1) WZNW-model,}
Nucl.\ Phys.\  B {\bf 792}, 257 (2008)
[arXiv:0708.0583 [hep-th]].

\bibitem{DI}
L.~Dolan and J.~N.~Ihry,
{\it Conformal supergravity tree amplitudes from open twistor string theory,}
Nucl.\ Phys.\  B {\bf 819}, 375 (2009)
[arXiv:0811.1341 [hep-th]].

\bibitem{Berezin:1987wh}
F.~A. Berezin, edited by A.A. Kirillov and D. Leites, 
{\it Introduction to
superanalysis}, Dordrecht, Netherlands: Reidel (1987) 
(Mathematical Physics and Applied Mathematics, vol. 9).


\end{thebibliography}
\end{document}